\documentclass[fleqn,usenatbib,useAMS]{mnras}

\usepackage{graphicx}	
\usepackage{amsmath}	
\usepackage{amssymb}	
\usepackage{multicol}        
\usepackage{bm}		
\usepackage{pdflscape}	

\usepackage[T1]{fontenc}
\usepackage{ae,aecompl}

\usepackage{newtxtext,newtxmath}

\title[WASP-7 differential rotation]{Can we detect the stellar differential rotation of WASP-7 through the Rossiter-McLaughlin observations?}

\author[L. M. Serrano]{Luisa Maria Serrano$^{1,2}$
    \thanks{Contact e-mail: \href{mailto:luisa.serrano@astro.up.pt}{luisa.serrano@astro.up.pt}}
    \thanks{Present address: Planetario do Porto, Rua das Estrelas, 4150-762, Porto, Portugal}, %
    M. Oshagh$^{1,3,6}$, %
    H. M. Cegla$^{4,5}$, %
    S. C. C. Barros$^{1}$, %
    N. C. Santos$^{1,2}$, 
    \newauthor
    J. P. Faria$^{1}$ and %
    B. Akinsanmi$^{1,2}$\\
$^{1}$Instituto de Astrof\'isica e Ci\^encias do Espa\c{c}o, Universidade do Porto, CAUP, Rua das Estrelas, PT4150-762 Porto, Portugal\\
$^{2}$Departamento de F\'isica e Astronomia, Faculdade de Ci\^encias, Universidade do Porto, Rua Campo Alegre, 4169-007 Porto, Portugal\\
$^{3}$Institut f\"ur Astrophysik, Georg-August-Universit\"at G\"ottingen, Friedrich-Hund-Platz 1, 37077, G\"ottingen, Germany\\
$^{4}$Observatoire de Gen\`eve, Universit\'{e} de Gen\`eve, 51 Chemin des Maillettes, Versoix 1290, Switzerland\\
$^{5}$CHEOPS Fellow, SNSF NCCR-PlanetS\\
$^{6}$Instituto de Astrof\'isica de Canarias (IAC), E-38200
La Laguna, Tenerife, Spain"\\}

\date{Last updated 2019 July 17}

\pubyear{2019?}

\begin{document}
\label{firstpage}
\pagerange{\pageref{firstpage}--\pageref{lastpage}}
\maketitle

\begin{abstract}
The Rossiter-McLaughlin (RM) effect is the radial velocity signal generated when an object transits a rotating star. Stars rotate differentially and this affects the shape and amplitude of this signal, on a level that can no longer be ignored with precise spectrographs. Highly misaligned planets provide a unique opportunity to probe stellar differential rotation via the RM effect, as they cross several stellar latitudes. In this sense, WASP-7, and its hot Jupiter with a projected misalignment of $\sim 90^{\circ}$, is one of the most promising targets. The aim of this work is to understand if the stellar differential rotation is measurable through the RM signal for systems with a geometry similar to WASP-7. In this sense, we use a modified version of \textit{SOAP3.0} to explore the main hurdles that prevented the precise determination of the differential rotation of WASP-7. We also investigate whether the adoption of the next generation spectrographs, like ESPRESSO, would solve these issues. Additionally, we assess how instrumental and stellar noise influence this effect and the derived geometry of the system. We found that, for WASP-7, the white noise represents an important hurdle in the detection of the stellar differential rotation, and that a precision of at least 2~m~s$^{-1}$ or better is essential.
\end{abstract}





\section{Introduction}
Stars rotate differentially and the stellar rotational period changes with the latitude \citep{Kitchatinov95,Kitchatinov99, Kuker, CollierCameron}. For the Sun, the rotational velocity decreases from the equator to the poles and many stars, mostly main sequence, show a solar-like differential rotation pattern \citep[and references therein]{Karoff}. For some cases, usually evolved stars, an anti-solar differential rotation was measured  \citep[][]{Kovari17, Kovari15, Harutyunyan}. Differential rotation is a key ingredient in stellar dynamo models, because it contributes significantly to the generation and maintanance of the stellar magnetic field \citep{Choudhuri}. The origin of stellar dynamos may be traced to the coupling between the stellar rotation, convection and conduction happening in the stellar plasma. The most widely accepted theory for the solar dynamo  states that the stellar differential rotation stretches the magnetic field lines, forcing it into a toroidal shape. This phenomenon is known as $\Omega$-effect. In a second moment, the field lines are twisted due to an effect known as $\alpha$-effect and they are unrolled, until the field returns to be poloidal. The overall stellar activity cycle strongly depends on the differential rotation pattern of the star \citep{Choudhuri}, though the underlying mechanisms which generate and sustain this non-rigid rotation are still poorly understood. Measuring this property for a large sample of stars (different spectral types and ages) provides a better understanding of the formation, and maintenance, of stellar magnetic activity and activity cycles as a whole. 
 
 Stellar differential rotation can be measured in several ways. Some of the most popular techniques require to measure period variations from long term photometric observations \citep{Reinhold} or through the analysis of chromospheric activity \citep{Donahue} or through spots detection in transit photometric and spectroscopic observations \citep{Zaleski, Morris17} or RV analysis \citep{Dumusque12}. Alternatively, we can follow the time migration of individual features on Doppler maps \citep{Donati, Barnes} or even study the effect of stellar differential rotation on line profiles \citep{Reiners00}. These methods allow one to measure the differential rotation for several stars, by estimating a parameter known as relative differential rotation, $\alpha$, i.e. the rotational sheer relative to the equator. The parameter $\alpha$ has been measured for several stars. For main sequence A-F types stars, no value above $0.45$ was retrieved through line profile analysis and the average results are between $0.1$ and $0.2$, with uncertainties between 0.07 and 0.30 \citep{Amlervoneiff}. For the same stellar types \citet{Balona} followed the time migration of spots on Doppler maps and reported an even lower average, $0.066$, with again similar errorbars. While this value is low, the range of $\alpha$ retrieved for A-F stars reaches 0.3, stressing for them the possibility of a stellar differential rotation stronger than the average. For G dwarfs, the authors reported higher mean values, between 0.072 and 0.126, with maximum $\alpha > 0.4$, while \citet{Karoff} used chromospheric analyses to retrieve $\alpha \sim 0.53$ for the sun-like star HD173701 (though, they do not report an uncertainty on this value). For the Sun, \citet{Berdyugina1} reports $\alpha \sim 0.2$. For K stars, \cite{Balona} reported an average value of $\alpha = 0.2$ and an upper limit of $0.4$, with errorbars again between 0.05 and 0.3; they also suggested that $\alpha$ has a higher average value for G-type stars and then decreases for earlier and later stellar types.
 
 The stellar differential rotation can affect the Rossiter-McLaughlin (RM) \citep{Holt, Rossiter, McLaughlin} effect, a phenomenon which happens because a planet transits a rotating star, blocking a portion of the stellar disc. The corresponding RV is removed from the integration of the radial velocity over the entire star. The RM signal allows to estimate the projected spin-orbit angle ($\lambda$), the angle between the normal vector of the orbital plane and the stellar rotational spin axis. So far, a wide diversity of $\lambda$ values have been measured, ranging from aligned \citep{Winn11} to highly misaligned systems \citep{Addison18}, and also retrograde planets \citep[e.g.][]{Hebrard}. Increasing the statistics on $\lambda$ is essential for testing theories on planet formation and evolution \citep[see e.g.][among the others]{Triaud, Albrecht12b, Winn10b}. For low mass planets, their orbits are generally aligned \citep{Fabrycky09}, while for hot Jupiters a stronger variety of $\lambda$ values has been observed. \citet{Albrecht12b} and \citet{Schlaufman10} show that the misalignments are stronger for hot Jupiters around hot stars ($T_{eff} > 6100 K$), while, in the case of very massive hot Jupiters ($M_P > 3 M_J$), the obliquities are lower \citep{Dawson14, Hebrard}. The physics driving these trends may be connected to star-planet tides, which tend to align the planetary orbits with the stellar equator. These tides should be stronger for cooler stars and for high massive planets. As mentioned in \citet{Mazeh15}, long period planets tend to have misaligned orbits (dissipation of tidal effects). Though, such theories are still debatable and increasing the statistics, accuracy and precision of $\lambda$ values will help to constrain these theories.
 
 A solar-like differential rotation changes the shape of the RM effect, decreasing the amplitude of the signal and affecting the estimate of $\lambda$. On the contrary an anti-solar like differential rotation will increase the amplitude of the RM signal. Thus, knowing $\alpha$ could allow a more accurate estimate of $\lambda$. Unfortunately, for the stars on which RM analysis is performed, the stellar differential rotation is not necessarily known, which poses the necessity of performing an estimation of $\alpha$. In this sense, the RM analysis can represent as well an alternative technique to measure the stellar differential rotation, allowing for an estimate of $\alpha$ and $\lambda$ at the same time. To explore this possibility, \citet{Gaudi07} adopted an analytic model of the RM signal which took into account the stellar differential rotation and they concluded that its effect was negligible in comparison to the RV precision of spectrographs at that time. Later, \citet{Hirano} showed that, for stars rotating faster than $v_* \sin i_* = $10~km~s$^{-1}$, the contribution of differential rotation in the RM signal could be crucial with the upcoming instruments. Since then, the spectrograph precision significantly improved, however there has been no conclusive determination of the stellar differential rotation through the RM signal. \citet{Cegla16} ruled out the possibility of rigid body stellar rotation for HD 189733, although they could not tightly constrain the level of $\alpha$ (which was found to be between $0.28$ and $0.86$). A similar result was obtained by \citet{Bourrier17} for the planet WASP-8b (they found $\alpha = 0.3 \pm 0.5$), though the authors could not conclude the effective presence of stellar differential rotation with certainty and considered the rigid model rotation the most plausible. \citet{Albrecht12} attempted a measurement of the stellar differential rotation for WASP-7. However, these authors speculated that systematic biases may have skewed their model fits towards unphysical solutions.

The present paper aims to assess which sources of noise prevented and still could hurdle the detection of stellar differential rotation through the RM signal and lead to an incorrect estimation of the system geometry. Note that we only explored solar-like differential rotation, since this seems to most likely happen for main sequence stars. Apart from the instrumental noise, the remaining sources of uncertainty have stellar origins. They are the center-to-limb variation of the convective blue-shift \citep{Dravins17, Cegla16, Shporer}, granulation, oscillations and the convective broadening due to granulation \citep[and references therein]{Kupka}. We then explored how well the stellar differential rotation could be characterized and the precision with which we can retrieve the projected spin-orbit angle.  We finally analyzed which instrumental characteristics can offer advantages when measuring the stellar differential rotation through RM analyses. In particular, we refer to the possibility of increasing the S/N ratio with a larger aperture telescope, and to use stable and precise spectrographs, such as HARPS \cite{Pepe} and HARPS-N, or new instruments, like ESPRESSO \citep{Pepe14} EXPRES \citep{Louis} and NEID \citep{Schwab}.

 In Sect. \ref{soap}, we report the updated version of \textit{SOAP3.0} \citep{Akinsanmi}, which takes into account the stellar differential rotation. In Sect. \ref{wasp-7b} we present our analysis of the observed RM signal of WASP-7b, reported by \cite{Albrecht12}. In Sect. \ref{simulations}, we present how we simulated the synthetic WASP-7b RM signal. In Sect. \ref{mainresults}, we explore the effect of several sources of noise on the detection of differential rotation. Finally, in Sect. \ref{discussion}, we identify the potential noise sources which prevented the detection of the stellar differential rotation in the past, and comment on the improvements offered by future instruments.

\section{\textit{SOAP3.0:} update with stellar differential rotation}
\label{soap}
As a first step, we updated the numerical tool \textit{SOAP3.0} \citep{Akinsanmi} by adding the stellar differential rotation as an extra feature. \textit{SOAP3.0}, a development of \textit{SOAP2} and \textit{SOAP-T} \citep{Oshagh1, Dumusque, Boisse}, describes the stellar surface as a grid and models the photometric transit light-curve and spectroscopic RM signal generated by a planetary transit in front of a rotating, potentially spotted stellar disk \footnote{Note that we switched off the simulations of stellar spots in our current study, assuming there are no magnetic activity features on the stellar surface.}. \textit{SOAP3.0} leaves to the user the choice to model the local cross-correlation function (CCF) of the stellar spectrum in two different ways: by adopting a solar CCF, as it was introduced in \textit{SOAP2.0}, or with a Gaussian, as it was already done in \textit{SOAP-T}. Each CCF (in a given stellar cell) can be modelled as a convolution of the instrumental and convective profiles (to account for convective broadening, i.e. macro-turbulence). We opted to use the Gaussian model because our work is based on an F star and for this stellar type no empirical local CCF is available yet. We are aware that convection could as well change the CCF profile, making it asymmetric. Not accounting for the correct shape could change the retrieved geometry of the system, as it was shown in \citet{Cegla}. The reloaded RM, \citep{Cegla16}, might be an alternative technique to isolate the CCF of F stars in the future. Nonetheless, even applying this method, \citet{Bourrier18}, \citet{Bourrier17} and \citet{Cegla16} showed that the local stellar CCF of G, K and M stars is well described by a Gaussian within their level of precision/SNR. When analyzing F stars with ESPRESSO, the Gaussian CCF should still be a good approximation, because we predict an instrumental noise similar to the HARPS values in the aforementioned papers. Nonetheless, a better modelling for the F star CCF could be performed within a future work.

For each grid cell in \textit{SOAP3.0}, the CCF is Doppler-shifted according to the projected rotational velocity. To account for the stellar differential rotation, we updated \textit{SOAP3.0}, and injected in an estimate of the local rotational velocity following the equation below, derived from solar observations:

\begin{equation}
\label{diff_rot}
\Omega(\theta) = \Omega_{\mathrm{eq}} (1 - \alpha \sin^2 \theta),
\end{equation}
where $\Omega(\theta)$ is the surface shear, defined as the angular velocity of rotation as a function of latitude. The parameter $\Omega_{\mathrm{eq}}$ is the equatorial angular velocity and $\theta$ is the stellar latitude \citep{Hirano, Cegla16}:
\begin{equation}
        \theta = \arcsin \left(z \sin i_* + \sqrt{1-\left(y^2+z^2\right)}\cos i_* \right)
\end{equation}
with $y,z$ the coordinates on the stellar disk and $i_*$ the inclination of the stellar rotational axis. When $i_* = 90$, the star is seen equator on and the latitude coincides with the position along the z axis. Finally, $\alpha$ is the relative differential rotation and it is defined as:
\begin{equation}
\alpha = \frac{\Delta\Omega}{\Omega_{\mathrm{eq}}},
\end{equation}
where $\Delta\Omega$ is the rotational frequency difference between the poles and the equator.

\section{Target selection}
As a first step we wanted to select a target with a high probability to detect the stellar differential rotation. To do so, we needed to identify a series of selection criteria, based on the intensity of the effect of a non rigid rotation on an RM signal. We thus performed a series of tests in which we explored under which conditions the effect of the stellar differential rotation became stronger on the observed RM. The amplitude of the RM signal strongly depends on several parameters, including the: planetary and stellar radii, orbital semi-major axis and inclination, stellar inclination, as well as the projected stellar rotation and projected spin-orbit angle. While the planetary radius can easily be deduced from a photometric transit analysis, determinations of the orbital inclination, stellar inclination, and projected spin-orbit angle may be impacted by the modeling (or lack thereof) of stellar differential rotation. To understand the impact these parameters may have on our determination of the stellar differential rotation, we performed a variety of tests. In the first test, we fixed all the parameters, except $\lambda$, which we varied between $0^{\circ}$, $30^{\circ}$, $60^{\circ}$ and $90^{\circ}$. In the second, we fixed $\lambda = 60^{\circ}$ and varied $i_P$, assigning values $90^{\circ}$, $89^{\circ}$, $88^{\circ}$, $87^{\circ}$ and $86^{\circ}$. In the third test, we fixed $i_P = 88^{\circ}$ and varied $i_*$, considering values $90^{\circ}$, $45^{\circ}$ and $30^{\circ}$. We report the details and results of these tests in Appendix \ref{Appendix1}. Inspecting the obtained results suggested that the strongest effect on RM happened for high projected spin-orbit angle, close to $90^{\circ}$, because in this condition the planet transits several latitudes, thus the effect of the stellar differential rotation is higher. On top of this, lower stellar inclination increases the impact of the stellar differential rotation on the observed RM, while orbital inclination does not seem to have a strong effect. Moreover, a higher projected rotational velocity will cause higher differences between RM signals with different values of $\alpha$.

Taking into account these effects, we selected a series of criteria for the target selection, as following:
\begin{itemize}
    \item projected spin-orbit angles in the intervals $45^{\circ}$ to $135^{\circ}$ and $-135^{\circ}$ to $-45^{\circ}$, to consider highly inclined systems;
    \item stars with high $v_* \sin i_*$, to target systems that will produce a larger RM signal. We limited this to the range $v_* \sin i_* = 5 - 40$~km~s$^{-1}$. Note that for very fast rotators the spectral lines are broadened, rendering the estimation of their parameters and RV analysis imprecise.
\end{itemize}
To such criteria we added two more:
\begin{itemize}
    \item stars with effective temperature, $5500$~K $ < T_{effe} < 7500$~K. The differential rotation depends on the stellar type and it is observed to be stronger for F-G stars \citep{Balona}.
    \item stars must have an RM signal already observed to be larger than 50~m~s$^{-1}$, to ensure possible RM observations for the future.
\end{itemize}
Note that we could not include the stellar inclination as additional parameter for the target selection, because it is usually unknown. Though, in Section \ref{diffrotation}, we perform additional tests to understand how freeing such parameters affects the RM fit. We applied these criteria to the projected spin-orbit angle catalog of TEPCAT \citep{Southworth11}. This selection criteria led us to isolate 5 systems CoRoT-01, WASP-7, WASP-79, WASP-100, WASP-109 and HAT-P-32b. We decided to select a specific target, WASP-7, for which there was already an attempt to measure its stellar differential rotation.

\section{WASP-7 and its hot Jupiter}
\label{wasp-7b}

\begin{table*}
\centering
\caption{Adopted parameters for simulating the RM of WASP-7 with \textit{SOAP3.0}. The stellar period of rotation was estimated from $v_* \sin i_*$, fixing $i_* = 90^{circ}$.}           
\label{simulationparameters}      
\begin{tabular}{ c c c c }
\hline\hline
& Stellar Properties & Value & Reference \\                   
\hline\hline
&&\\
$R_*  \; [R_{\odot}]$ &  stellar radius & $1.432 \pm 0.09$ & \citet{Southworth11}\\
$T_{\mathrm{eff}} \; [K]$ & effective temperature & $6520 \pm 70$ & \citet{Maxted11}\\
$P_*$ [~days] & stellar rotational period & $5.18$\\
$v_* \sin i_*$ [~km~s$^{-1}$] & projected rotational velocity & $14 \pm 2$ & \citet{Albrecht12}\\
$i_*$ [$^{\circ}$] & stellar axis inclination &  $90$ (fixed) & \citet{Albrecht12}\\
$u_1$ & linear limb darkening &  $0.2$ (fixed)\\
$u_2$ & quadratic limb darkening &  $0.3$ (fixed)\\
\hline\hline

& Planetary Properties  &  Value & Reference\\                   
\hline\hline
&&\\

$R_p \; [R_J]$ & planet radius &  $1.33 \pm 0.093$ & \citet{Southworth11}\\
$P_p$[~days] & planet orbital period & $4.9546416 \pm 3.5 \times 10^{-6}$ & \citet{Southworth11}\\
$a \; [AU]$ & semi-major axis  & $0.0617 \pm 0.0011$ & \citet{Southworth11}\\
$i_p$ [$^{\circ}$] & orbital inclination & $87.03 \pm 0.93$ & \citet{Southworth11}\\
$\lambda$ [$^{\circ}$] & projected spin-orbit angle & $86 \pm 6$ & \citet{Albrecht12}\\

\hline
\end{tabular}
\end{table*}

\begin{figure}
\centering
\includegraphics[width=9.5cm]{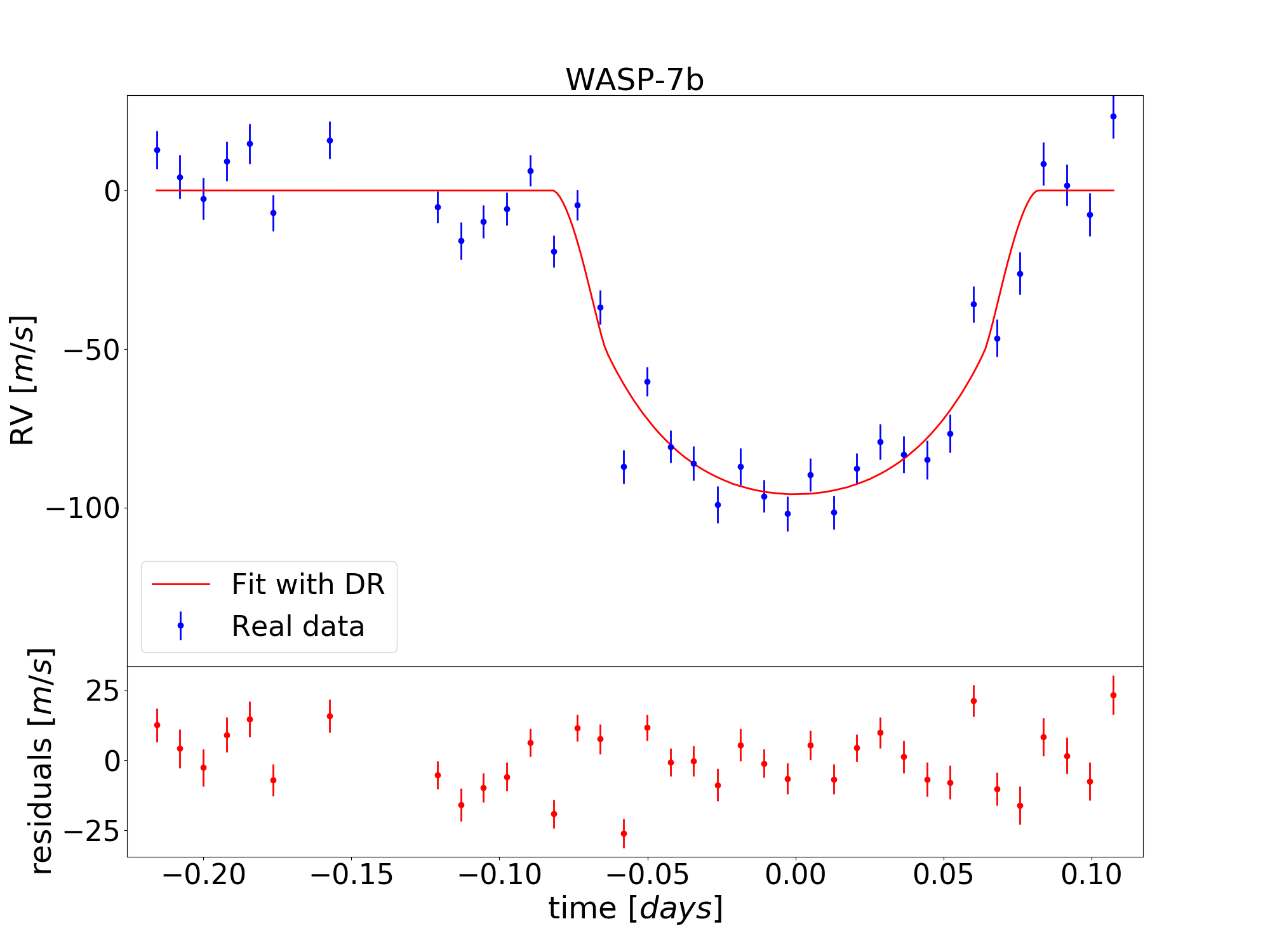}
\caption{Fit of the PFS data of WASP-7b using the updated \textit{SOAP3.0}
with differential rotation (DR). Top: the blue error bars represent the observed data by \citet{Albrecht12} 
for WASP-7b, while the thick red line is the best fit reported in the present paper for the RM signal. Bottom: residuals of the observed RM with respect to the best fit. }
\label{RealtestWASP7}
\end{figure}

When the projected spin-orbit angle is close to $90^{\circ}$, it is more likely that a transiting planet occults many stellar latitudes, which makes it easier to determine the latitudinal dependence of the stellar differential rotation. Due to a projected spin-orbit angle nearly perpendicular to our line-of-sight, the hot Jupiter WASP-7b is one of the most promising target for measuring $\alpha$ (see Table \ref{simulationparameters} for the main properties of the system). It orbits an F5V-type star with apparent visual magnitude 9.51. \citet{Albrecht12} observed the RM signal with the Planet Finder Spectrograph (PFS), which has a resolution of $R = 38000$ (improvable until $190000$) and a potential instrumental precision of 1.25~m~s~$^{-1}$ \citep{Crane}. However, this instrumental precision decreases for both dimmer and faster rotating stars; for WASP-7, the observed RV precision for an $10$~minutes (min.) exposure is 5.6~m~s~$^{-1}$. We refer to \cite{Albrecht12} for more details on the quality of the data. \cite{Albrecht12} performed a $\chi^2$ minimization on the RM, estimating a projected spin-orbit angle $\lambda_0 = 86^{\circ} \pm 6^{\circ}$ and projected rotational velocity $(v_* \sin i_*)_0= 14.0 \pm 2.0$~km~s~$^{-1}$ \citep{Albrecht12}. The authors found $\alpha_0 = 0.45 \pm 0.1$ when the prior on $\alpha$ was restricted to $0.1 - 0.5$. Then, they significantly expanded the $\alpha$ prior and retrieved $\alpha = 0.9$ (no error was reported in this case), which would indicate that the pole rotates extremely slower with respect to the equator. This condition is very extreme and has never been observed so far, especially for F stars \citep[see][]{Reiners1, Balona}, so they expressed doubts on the values they obtained for $\alpha$. 

We analyzed the WASP-7 data from \cite{Albrecht12} with the updated \textit{SOAP3.0}. We first fitted a linear polynomial to the out-of-transit data to remove the effect of the Keplerian orbit and non-occulted star-spots from the RV data. We then performed a $\chi ^2$ minimization on the RM signal using the updated \textit{SOAP3.0} model. In our fit, we used the resolution of the PFS and as FWHM of the CCF $10.3$~km~s$^{-1}$, which accounts for both the instrumental broadening and the convective broadening \citep[see][for details]{Doyle}. As limb darkening parameters, we approximated those reported in \cite{Claret} for F-type stars ($u_1 = 0.2$ and $u_2 = 0.3$), considering that \cite{Albrecht12} expressed doubts on the coefficients they estimated ($u_1 = 0.17$ and $u_2 = 0.38$). As mentioned in \cite{Hirano}, accounting for the stellar differential rotation allows to break the degeneracy between the stellar period of rotation and the stellar inclination. Fitting both $P_*$ and $i_*$ can be reasonable after $\alpha$ is well constrained. For $i_* \neq 90^{\circ}$, the amplitude of the RM decreases, because the planet transits a lower number of latitudes and, in the case of WASP-7, it will cross areas of the star which are rotating more slowly. This will degrade the precision with which $\alpha$ is retrieved. Nonetheless, the objective of our work is to verify if we can measure $\alpha$. For this reason, we used $i_* = 90^{\circ}$, as \citet{Albrecht12} estimated using the method by \citet{Schlaufman10}. In the fit, we fixed all the parameters to the values in Table~\ref{simulationparameters}, except for $v_* \sin i_*$, $\lambda$ and $\alpha$. We varied the free parameters in the intervals: $\lambda$ from $61^{\circ}$ to $111^{\circ}$ with a step of $0.4^{\circ}$, $v_* \sin i_*$ from $8$ to $20$~km~s$^{-1}$ with a step of $0.25$~m~s$^{-1}$, and $\alpha$ between $0$ and $1$, with a $0.033$ step. These steps were chosen because we saw that the fit remained unchanged if they were smaller. Throughout the work, we report the $1 \sigma$ uncertainties on the best fit values. When the error-bar was large enough to go beyond the explored range of values for $\alpha$ ($[0,1]$), we imposed as errors the differences of the best fit with respect to the upper or lower limit.

To test the reliability of the $\chi^2$ minimization, we used it to fit a mock simulation of a WASP-7b RM with $\alpha = 0.6$ and without instrumental noise. The best-fit parameters were exactly equal to the injected ones.

The fit on real data gave $\alpha = 0.75^{+0.25}_{-0.41}$, $v_* \sin i_* = 14.5 \pm 1.0$~km~s$^{-1}$ and $\lambda = 90^{\circ} \pm 4^{\circ}$. This result is compatible with that of \cite{Albrecht12} within $1\sigma$. In contrary to them, we find smaller error-bars on the $v_*\sin i_*$ and $\lambda$ and a larger error on $\alpha$. In Figure \ref{RealtestWASP7}, we report the observed data and our best fitted model. In the lower frame, we also show the residuals, which have a maximum amplitude of 25~m~s$^{-1}$ but an RMS of 11~m~s$^{-1}$, which is higher than the instrumental noise of 5.6~m~s$^{-1}$. This suggests that the presence of an extra unexplained noise of 9.8~m~s$^{-1}$. At $3\sigma$, we cannot constrain $\alpha$ and the estimated result is higher than what is expected and has been observed for stars of this type.

\section{Simulations}
\label{simulations}
In this section, we present how we modelled the mock RM signals which allowed us to test detectability of the stellar differential rotation through RM observations. We produced simulations of WASP-7b RM signal with a $6$~min exposure time and spectral resolution $R = 140000$, as offered by the spectrograph ESPRESSO. This instrument promises to achieve a precision of 10~cm~s$^{-1}$ for bright stars. For the RM analysis its greatest advantage comes from the collective power of the 8~m aperture Very Large Telescope (VLT). This allows us to obtain more data points at a given signal-to-noise ratio than a 4~m class telescope, when sampling the RM of bright stars. To have a first estimate of the ESPRESSO RV precision with good atmospheric conditions, we used the ESPRESSO Exposure Time Calculator (ETC), which performs estimates for G, K and M stars. To produce a more realistic simulation, we modelled the mock RM signals assuming a local CCF FWHM $6.4$~km~s$^{-1}$. This width, defined as the sum in quadrature between the instrumental and convective broadening, is smaller than the one applied to fit the PFS observations, because ESPRESSO instrumental FWHM, 2.067~km~s$^{-1}$ (as estimated from the definition $FWHM = c/R$, with $c$ the speed of light), is lower. All simulations include stellar differential rotation and instrumental noise as described in Section~\ref{WN}. On top of this, we also explored the additional impact of specific stellar noise sources, the center-to-limb variation of the convective blue-shift, in Section \ref{CLVCB}, a red noise source, i.e. granulation, and oscillations in Section \ref{GRANOSCILL}. We did not include other red noise sources such as instrumental systematic, since they are not well-known and they are not reproducible.

\subsection{Instrumental noise}
\label{WN}
We modelled the instrumental noise as a white noise with standard deviation equal to the precision of the instrument. We considered several levels of white noise. The first two noise levels, $2$~m~s$^{-1}$ and $3$~m~s$^{-1}$, are determined considering the stellar type of WASP-7, which is an F star. We estimated these values following the method described in \citet{Bouchy}. The instrumental noise depends on the efficiency of the spectrograph, telescope aperture, stellar type and rotational velocity of the star, $v_*\sin i_*$. It is also possible there might be a dependence on the spectrograph resolution as well. We tested whether the retrieved $v_*\sin i_*$, $\lambda$ and $\alpha$ were more precise with ESPRESSO resolution if compared to HARPS resolution and this was not the case. To estimate the instrumental noise, we used the ESPRESSO ETC to retrieve the predicted value for a K star with the same magnitude as WASP-7, $0.32$~m~s$^{-1}$. In this way, it was already possible to account for the telescope aperture and the instrumental efficiency. We then scaled this value, accounting for the quality factors of a non rotating K star and of a rotating F star (reported in Table~2 in \cite{Bouchy}). Note that, when choosing a K star to estimate the final instrumental noise of an F star, we obtain the same instrumental noise we would get selecting at first a G star and applying its quality factor. For averagely rotating F stars (8~km~s$^{-1}$), we estimated $2$~m~s$^{-1}$, while for a star as fast as WASP-7 we obtained $3$~m~s$^{-1}$. Besides these, we also considered three lower noise levels, $0.5$, $1$ and $1.5$~m~s$^{-1}$, to test the detectability of the stellar differential rotation if the star belonged to other stellar types, such as G and K. Finally, we also tested other scenarios with increased instrumental noise of $5$~m~s$^{-1}$ and $7$~m~s$^{-1}$. An instrumental noise of $5$~m~s$^{-1}$ corresponds the predicted precision for HARPS with a $6$~min exposure time for a star similar to WASP-7 \citep[e.g.][]{Hellier18}, while $7$~m~s$^{-1}$ is the noise for the PFS spectrograph and other less precise instruments. 
 
 \subsection{Center-to-limb variation of the convective blue-shift}
 \label{CLVCB}
The convective blue-shift (CB) is a consequence of granulation and it occurs because the emerging granules are brighter and cover a greater fraction of the stellar surface with respect to the inter-granular lanes. It results in an additional blue-shift between the measured stellar line positions and their laboratory counterparts \citep{Adam}. \citet{Shporer} modelled the CB, considering it as a constant effect, which changed across the center-to-limb angle due to limb darkening and the projected area. They argued that ignoring the CB should influence the estimation of the projected spin-orbit alignment. \cite{Cegla} improved on this by adopting a 3D magneto-hydrodynamic (MHD) solar simulation to determine the center-to-limb variation of the CB effect. They also included the impact of an asymmetric line profile on the stellar disc; this is significant as the granulation is known to introduce asymmetries. In particular, \cite{Cegla} report that ignoring these two convective effects on moderately rotating stars (e.g. $v_* \sin i_* = 6$~km~s$^{-1}$) could potentially inject systematic biases of $\sim 20^{\circ}$ or more in the projected obliquities. An updated version of their MHD simulation, and corresponding CB predictions, was presented in \cite{Cegla18}. To account for the center-to-limb CB, we shifted the local radial velocity of \textit{SOAP3.0}, adding to it a fourth order polynomial valid for a solar CB with a net magnetic field of 0\,G \citep{Cegla18}.

\subsection{Granulation and pressure-mode oscillations}
\label{GRANOSCILL}
Granulation is a red noise \citep{Hekker17} caused by convective patterns at the stellar surface, happening over a few minutes and up to $\sim$1-2~days for main sequence stars \citep{DelMoro,Christensen-Dalsgaard}. The oscillations are a stellar photospheric phenomenon related to the propagation of pressure waves at the surface of solar-type stars, excited by convection. The p-mode oscillations lead to a motion of the external stellar envelope over timescales of a few minutes for main sequence stars.

These noise sources can introduce an uncertainty in the detection of stellar differential rotation. To perform a first simple test, we modelled these noise sources, by generating synthetic RV measurements with the components proposed by \citet{Dumusque11}. We considered Harvey-like functions to model granulation,
mesogranulation, and supergranulation \citep{Harvey} and a Lorentzian function to model the frequency bump due to oscillations \citep[e.g.]{Lefebvre}. As time scales and amplitudes we used the values determined for $\beta$ Hyi \citep[see Table 2 in]{Dumusque11} as it has the closest spectral type to WASP-7. The model for the power spectral density is finally inverted to obtain RV values at the simulated times of the RM signal. In detail, we produced two RV sets, one only accounting for granulation, and the second one with both granulation and oscillations.

\section{Results}
\label{mainresults}
In this section, we present all the tests we performed on mock RMs of WASP-7b and the main results we obtained.

\subsection{Minimum detectable $\alpha$}
\label{diffrotation}
\begin{table*}
\caption{Results of our fitting procedure for the WASP-7 RM simulations,} including instrumental noise ($\sigma = 2$~m~s$^{-1}$) and differential rotation. 0n the left side, we report the results of the fit performed accounting for rigid rotation in the model, while on the right we show the results obtained as we inject the stellar differential rotation in the fitting model. The input $v_*\sin{i_*}$, $\lambda$ and $\psi$ were 14~km~s$^{-1}$, $86^{\circ}$ and $85.21^{\circ}$.             
\label{resultsvariousalpha}      
\centering                          
\begin{tabular}{ c c c c c c c c c c }   
\hline\hline
Simulation && \multicolumn{2}{c}{Fit with $\alpha = 0$} && \multicolumn{3}{c}{Fit varying $\alpha$}\\
\hline\hline
input $\alpha$ && $v_*\sin i_*$ & $\lambda$ & $\psi$ && $v_*\sin i_*$ & $\lambda$ & $\alpha$ & $\psi$ \\  
\hline 
$0.1$ && $13.8 \pm 0.4$~km~s$^{-1}$ & $84.4^{\circ} \pm 0.8^{\circ}$ & $84.41^{\circ} \pm 0.80^{\circ}$ && $14.0 \pm 0.5$~km~s$^{-1}$ & $86.0^{\circ} \pm 1.2^{\circ}$ & $0.13^{+0.20}_{-0.13}$ & $86.01^{\circ} \pm 1.20^{\circ}$\\
$0.2$ && $13.5 \pm 0.3$~km~s$^{-1}$ & $84.6^{\circ} \pm 1.5^{\circ}$ & $84.61^{\circ} \pm 1.50^{\circ}$ && $14.0 \pm 0.3$~km~s$^{-1}$ & $86.0^{\circ} \pm 1.8^{\circ}$ & $0.23 \pm 0.04$ & $86.01^{\circ} \pm 1.80^{\circ}$\\
$0.3$ && $13.0 \pm 0.3$~km~s$^{-1}$ & $84.2^{\circ} \pm 0.8^{\circ}$ & $84.21^{\circ} \pm 0.80^{\circ}$ && $14.0 \pm 0.3$~km~s$^{-1}$ & $85.2^{\circ} \pm 1.2^{\circ}$ & $0.30 \pm 0.10$ & $85.21^{\circ} \pm 1.20^{\circ}$\\
$0.4$ && $12.8 \pm 0.3 $~km~s$^{-1}$ & $84.2^{\circ} \pm 1.1^{\circ}$ & $84.21^{\circ} \pm 1.10^{\circ}$ && $14.0 \pm 0.8$~km~s$^{-1}$ & $86.0^{\circ} \pm 1.6^{\circ}$ & $0.40 \pm 0.05$ & $86.01^{\circ} \pm 1.60^{\circ}$\\
$0.5$ && $12.8 \pm 0.3$~km~s$^{-1}$ & $81.3^{\circ} \pm 0.6^{\circ}$ & $81.31^{\circ} \pm 0.60^{\circ}$ && $14.0 \pm 0.8$~km~s$^{-1}$ & $86.0^{\circ} \pm 1.8^{\circ}$ & $0.50 \pm 0.10$ & $86.01^{\circ} \pm 1.80^{\circ}$\\
$0.6$ && $12.5 \pm 0.3$~km~s$^{-1}$ & $80.7^{\circ} \pm 1.2^{\circ}$ & $80.71^{\circ} \pm 1.20^{\circ}$ && $14.0 \pm 0.5$~km~s$^{-1}$ & $85.2^{\circ} \pm 1.6^{\circ}$ & $0.60 \pm 0.10$ & $85.21^{\circ} \pm 1.60^{\circ}$\\

\hline
                               
\end{tabular}
\end{table*}
\begin{table*}
\caption{Results of the fitting procedure with 4 free parameters for the WASP-7 rm simulations, with various $i_*$ and including instrumental noise ($\sigma = 2$~m~s$^{-1}$) and differential rotation. The 4 free parameters were $i_*$, $P_*$, $\lambda$ and $\alpha$. At the top, we report the results of the fit for $\alpha = 0.3$, while at the bottom for $\alpha = 0.6$.}             
\label{Varying_stellar_inclination}      
\centering                          
\begin{tabular}{ c c c c c c c c  }   
\hline\hline
\multicolumn{8}{c}{$\alpha = 0.3$} \\
\hline\hline
input $i_*$ & input $P_*$ & input $\psi$ & $i_*$ & $P_*$ & $\lambda$ & $\alpha$ & $\psi$ \\ 
\hline 
$90^{\circ}$ & $5.18$~days & $86.01^{\circ}$ & $91.0^{\circ} \pm 1.0^{\circ}$ & $5.18 \pm 0.15$ days & $86.0^{\circ} \pm 1.2^{\circ}$ & $\alpha = 0.32 \pm 0.11$ & $86.06^{\circ} \pm 1.20^{\circ}$ \\
$60^{\circ}$ & $4.48$~days & $86.05^{\circ}$ & $60.5^{\circ} \pm 1.5^{\circ}$ & $4.48 \pm 0.15$ days & $85.2^{\circ} \pm 1.6^{\circ}$ & $\alpha = 0.32 \pm 0.11$ & $84.36^{\circ} \pm 1.47^{\circ}$ \\
$45^{\circ}$ & $3.66$~days & $86.07^{\circ}$ & $46.5^{\circ} \pm 1.5^{\circ}$ & $3.66 \pm 0.15$ days & $84.4^{\circ} \pm 1.8^{\circ}$ & $\alpha = 0.37 \pm 0.13$ & $83.89^{\circ} \pm 1.45^{\circ}$ \\

\hline\hline 
\multicolumn{8}{c}{$\alpha = 0.6$}\\
\hline\hline
input $i_*$ & input $P_*$ & $\psi$ & $i_*$ & $P_*$ & $\lambda$ & $\alpha$ & $\psi$ \\ 
\hline\hline                
$90^{\circ}$ & $5.18$~days & $86.01^{\circ}$ & $88.5^{\circ} \pm 1.5^{\circ}$ & $5.18 \pm 0.15$ days & $85.2^{\circ} \pm 1.8^{\circ}$ & $\alpha = 0.62 \pm 0.05$ & $85.13^{\circ} \pm 1.80^{\circ}$\\
$60^{\circ}$ & $4.48$~days & $86.05^{\circ}$& $60.5^{\circ} \pm 1.5^{\circ}$ & $4.48 \pm 0.15$ days & $85.2^{\circ} \pm 1.2^{\circ}$ & $\alpha = 0.62 \pm 0.11$ & $84.36^{\circ} \pm 1.14^{\circ}$\\
$45^{\circ}$ & $3.66$~days & $86.07^{\circ}$ & $46.0^{\circ} \pm 1.5^{\circ}$ & $3.66 \pm 0.15$ days & $85.2^{\circ} \pm 1.6^{\circ}$ & $\alpha = 0.64 \pm 0.16$ & $84.48^{\circ} \pm 1.32^{\circ}$\\
       \hline                        
\end{tabular}
\end{table*}

\begin{figure}
\centering
\includegraphics[width=9.5cm]{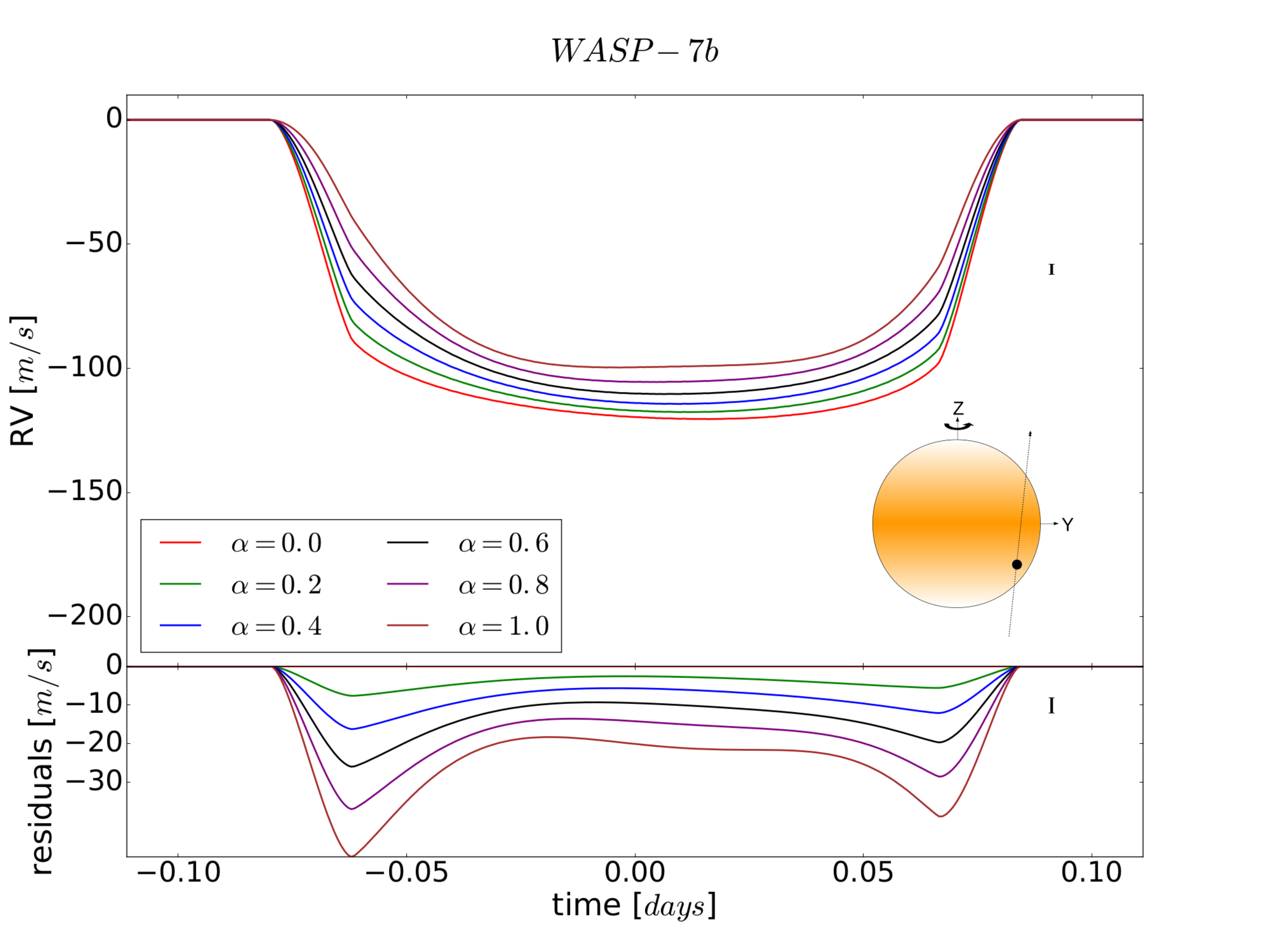}
\caption{Top: RM simulations for the planet WASP-7b and six different values of $\alpha$, the relative differential rotation, $0.0$, $0.2$, $0.4$, $0.6$, $0.8$ and $1$. Bottom: residuals of the RM simulation in the top plot with respect to the model without differential rotation ($\alpha = 0$). The vertical black lines in the right side of the two frames represent the ESPRESSO error for averagely fast rotating F stars, which is 2~m~s$^{-1}$, and they are added to allow a visual comparison with the effect of the differential rotation on RM. In the blank area of the top frame we also show a schematic geometry of the system. The stellar disk is represented as an orange disk. As the latitude increases, the orange fades to white to give an idea of how the rotational velocity decreases.}
\label{WASP-7b}
\end{figure}
\begin{figure*}
\centering
\includegraphics[width=8.7cm]{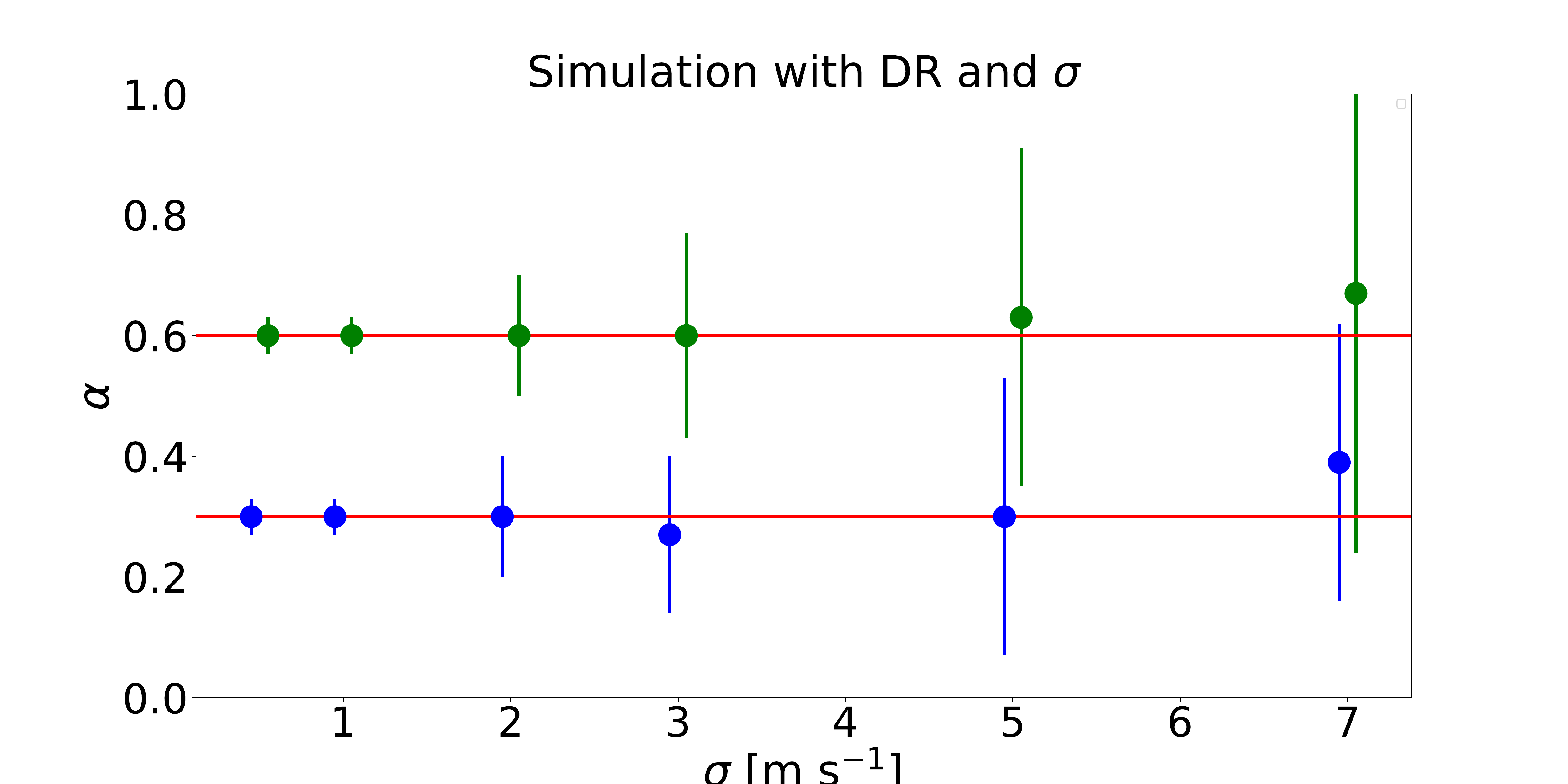}
\includegraphics[width=8.7cm]{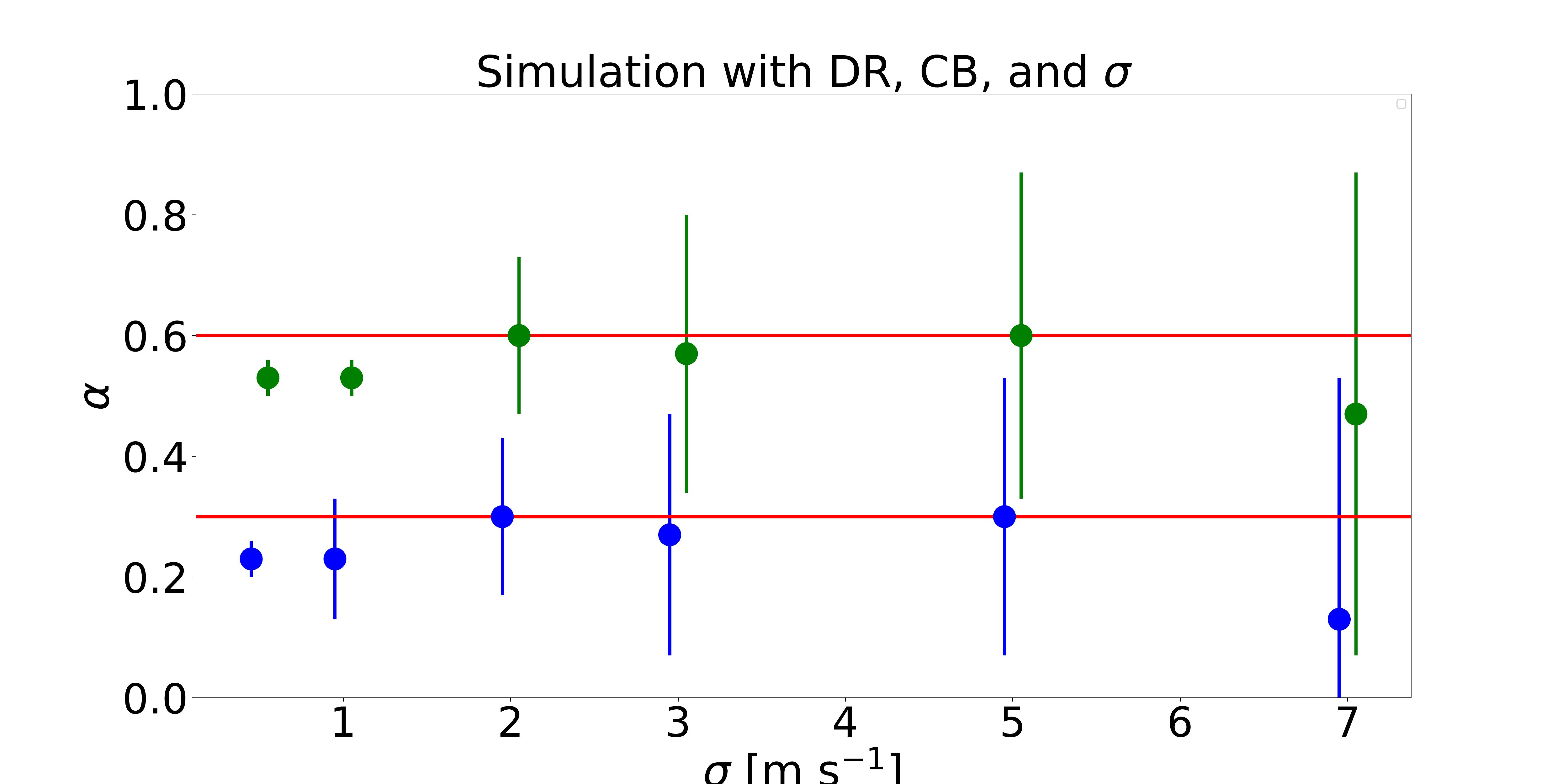}
\includegraphics[width=8.7cm]{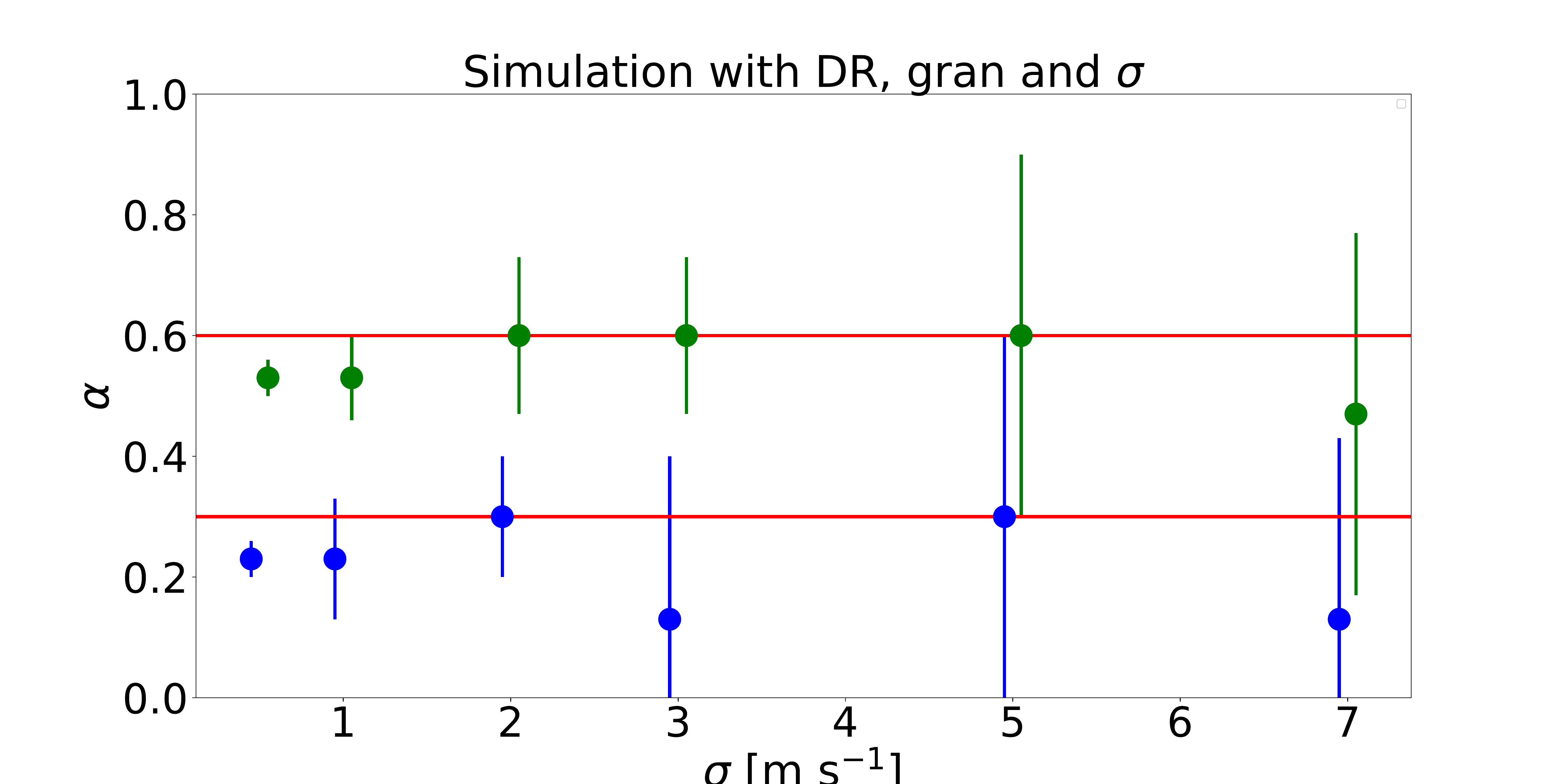}
\includegraphics[width=8.7cm]{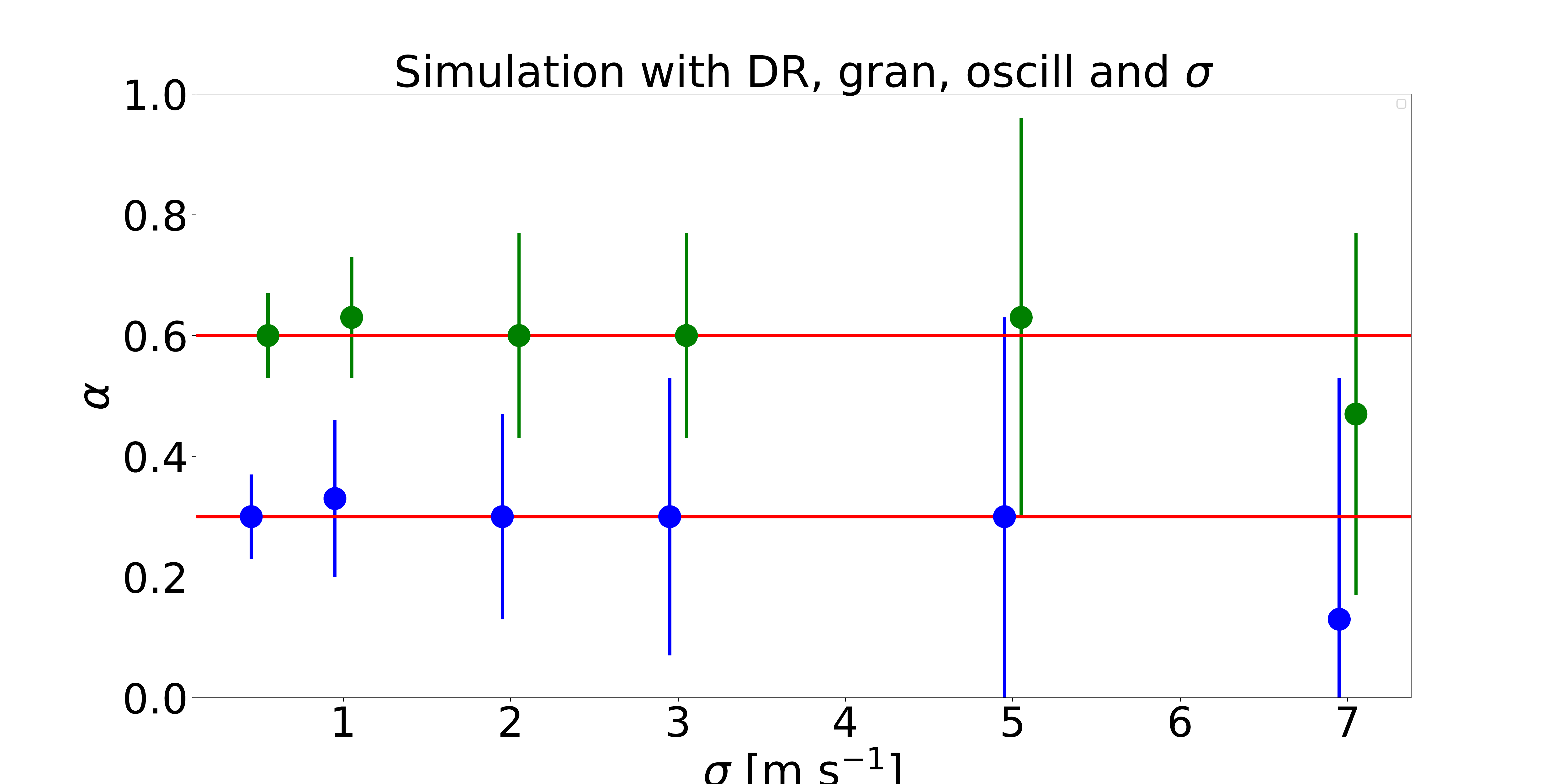}
\caption{Best fit $\alpha$ and relative error-bars as a function of the instrumental noise. Top left: results for simulations of RM which included differential rotation (DR) and instrumental noise ($\sigma$). Top right: results for simulations also with center-to-limb variation of the convective blue-shift (CB). Bottom left: results for simulations also with granulation (gran). Bottom right: results for simulations including the oscillations (oscill) too. The blue errorbars are the best fits for input $\alpha = 0.3$, the green ones for $\alpha = 0.6$. The fit accounts only for the differential rotation in the model.}
\label{Results}
\end{figure*}

To visualize how the differential rotation alone affects the RM signal of a planet like WASP-7b, in Figure~\ref{WASP-7b}, we show different models of WASP-7b for $\alpha$ from $0.0$ to $1.0$. In the bottom, we represent the residuals of each model with respect to the rigid rotation case ($\alpha = 0$). We also add a schematic view of the system inside the plot. This way we show which areas of the stellar disc are transited by the planet. Since the projected spin-orbit angle is close to polar configuration, the planet transits several latitudes, so the stellar differential rotation significantly affects the RM signal. Inspecting the mock RMs in Figure~\ref{WASP-7b}, we see that the effect of differential rotation is stronger on the transit ingress and egress and it decreases as we get closer to the mid-transit time. At mid-transit, the difference between $\alpha = 0$ and $1$ is 20~m~s$^{-1}$, whereas between $\alpha = 0$ and $\alpha = 0.6$, it is $10$~m~s$^{-1}$. Our objective is to verify if this difference is detectable. From now on, we will consider as maximum value of $\alpha = 0.6$, because a stronger stellar differential rotation is unlikely and the average $\alpha$ for F type stars is much lower, around $0.1-0.2$.

We produced RM simulations which included a noise level of $2$~m~s$^{-1}$ and various $\alpha$ values, $0.1$, $0.2$, $0.3$, $0.4$, $0.5$ and $0.6$. We fitted them twice, once assuming rigid rotation and the second time accounting for the stellar differential rotation. The results are in Table \ref{resultsvariousalpha}, on the left panel are the tests with $\alpha$ fixed to $0$ in the fitting model, while on the right $\alpha$ is a free parameter of the $\chi^2$. In this table we also report the uncertainty when deriving the spin-orbit angle $\psi$. To estimate $\psi$ we used:
\begin{equation}
    \cos \psi = \cos i_* \cos i_p + \sin i_* \sin i_p \cos \lambda
\end{equation}
and propagated the errors accordingly.

When we assume rigid rotation, as $\alpha$ increases, $v_* \sin i_*$ and $\lambda$ estimates become more inaccurate and less precise with respect to the input values, $\lambda_0 = 86^{\circ}$ and $(v_* \sin i_*)_0 = 14$~km~s$^{-1}$. As $\alpha > 0.3$, both the recovered $v_*\sin i_*$ and $\lambda$ become incompatible with their true input values by at least $2 \sigma$. For the extreme case of $\alpha = 0.6$, $\lambda$ diverges from the input value by $5.3^{\circ}$, while the $v_*\sin i_*$ diverges by $1.5$~km~s$^{-1}$. For lower $\alpha$, the differences between the best fit and the input values are smaller; however, the two values are incompatible within the uncertainties. On the other hand, as we see in the right panel of Table~\ref{resultsvariousalpha}, when $\alpha$ is a free parameter and the input $\alpha \ge 0.2$, all of the parameters are retrieved. For smaller input $\alpha$, our recovery could not rule out rigid body rotation within $1\sigma$. For input $\alpha$ < 0.2, a star with rigid rotation can fit the data just as well as that with differential rotation. We also note that the retrieved $\psi$ is compatible with the injected one of $86.01^{\circ}$ well within $1 \sigma$ .

To test what would happen whether we consider the stellar inclination as a free parameter, we decided to perform an additional test. We assumed three values of the stellar inclination, $i_* = 90^{\circ}$, $60^{\circ}$ and $45^{\circ}$. Decreasing $i_*$ reduces the amplitude of the RM signal (because the planet transits regions of the star with lower rotation velocity). To keep the same amplitude as the previous tests, in compatibility with the observed data, we varied as well the period of rotation of the star at the equator, to compensate the effect of the stellar inclination. As a result for $i_* = 90^{\circ}$ we kept $P_* = 5.18$~days, for $i_* = 60^{\circ}$ we imposed $P_* = 4.48$~days, while for $i_* = 45^{\circ}$ we used $P_* = 3.66$~days. In this way, we produced 6 additional mock simulations of WASP-7, adding to each of them a white noise on the level of 2~m~s$^{-1}$. We performed our fit, using as free parameters $i_*$, $\lambda$, $P_*$ and $\alpha$ and adopting as model the updated \textit{SOAP3.0}. We report our results in Table \ref{Varying_stellar_inclination}. At the top, we list the best fit parameters for the case with $\alpha = 0.3$, while at the bottom those for $\alpha = 0.6$. We also show how varying the stellar inclination affects the estimated $\psi$. The main result is that the best fit parameters are compatible within $1\sigma$ with the input values, though the retrieved $\alpha$ tends to increasingly differ with respect to the input as $i_*$ decreases. This effect is more evident in the case of $\alpha = 0.3$. The stellar rotational velocity is estimated with high precision, allowing us to break the degeneracy between $i_*$ and $P_*$.

These results show that if the differential rotation is not accounted for in the fitting model, we obtain systematically biased estimates of the obliquity of the system. Moreover, without including stellar noise in the mock RMs, when the instrumental noise is $2$~m~s$^{-1}$, the lowest measurable $\alpha$ for a planetary system like WASP-7 is $0.2$. Additionally we demonstrate that the stellar differential rotation allows us to derive $i_*$ through the RM signal.

\subsection{Varying the instrumental noise}
\label{varying_instrumental_noise}
\begin{figure}
\centering
\includegraphics[width=9.5cm]{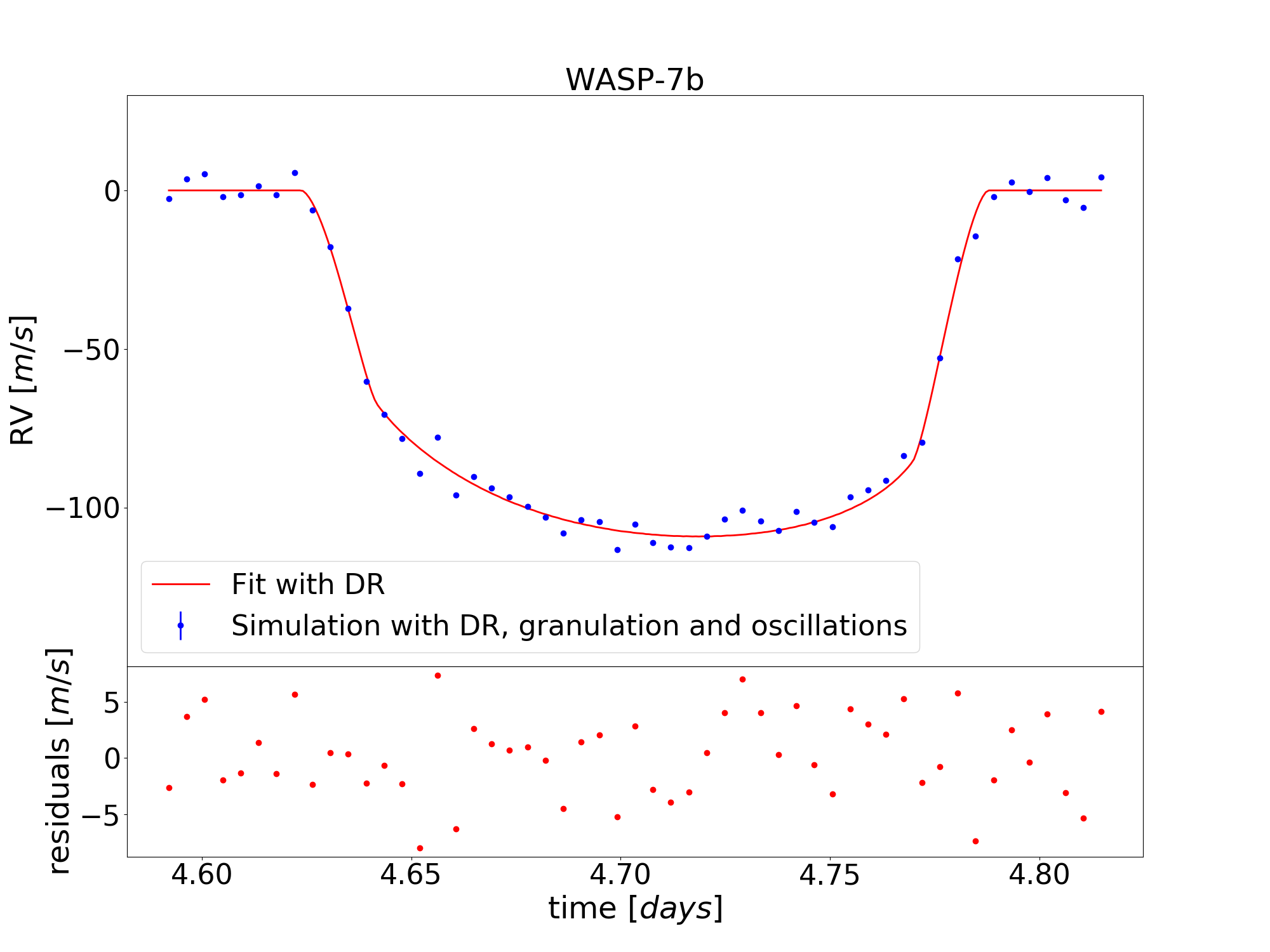}
\caption{Fit of the mock data of WASP-7b which include differential rotation (DR) $\alpha = 0.3$, granulation (gran) oscillation (oscill) and a white noise of 2~m~s$^{-1}$. The fitting model accounts only for differential rotation. Top: the blue error-bars represent the simulated data, while the thick red line is the best fit. Bottom: residuals of the modelled RM with respect to the best fit.}
\label{Testgranosc}
\end{figure}

As a following step, we performed four different series of tests as follows. We produced simulations of WASP-7b RM with $\alpha = 0.3$ and $0.6$. The case of $\alpha = 0.6$ represents an upper limit based on the literature. On the other hand, choosing $0.3$ was justified by the large uncertainty retrieved for lower $\alpha$ in the tests only with instrumental noise (for instance, $0.2 \pm 0.13$). Adding extra noise would render the detection less reliable. We then added different levels of white noise to test the detectability of $\alpha$. In the first series of tests, no other noise source was included; in the second we accounted for the solar 0G center-to-limb CB in the mock RMs (see Section \ref{CLVCB}); in the third we replaced the CB with granulation (as estimated for $\beta$ Hyi, see Section \ref{GRANOSCILL}); while in the last we included both granulation and oscillations in the simulated data, since they are usually coupled. This way we could analyze how each of the mentioned phenomena can influence the estimate of $\alpha$ as the instrumental precision varies.

The best fit values for the three free parameters, $\lambda$, $\alpha$ and $v_* \sin i_*$ are reported in Tables~\ref{resultsRMwithDR0.3}, \ref{resultsRMwithDR0.6}, \ref{resultsRMwithDRCBGRANOSC0.3} and \ref{resultsRMwithDRCBGRANOSC0.6}. In particular, in Tables~\ref{resultsRMwithDR0.3} and \ref{resultsRMwithDR0.6}, we compare the best fits of the simulations only with instrumental noise (in the left side) with respect to those of the mock data also with CB. In the left panels of Tables~\ref{resultsRMwithDRCBGRANOSC0.3} and \ref{resultsRMwithDRCBGRANOSC0.6}, we show the results for the simulations only with granulation, while in the right we report the retrieved parameters for mock data with granulation and oscillations.

In Figure~\ref{Results}, we show the results reported in Tables~\ref{resultsRMwithDR0.3}, \ref{resultsRMwithDR0.6}, \ref{resultsRMwithDRCBGRANOSC0.3} and \ref{resultsRMwithDRCBGRANOSC0.6}. In particular, we report the dependence of the recovered $\alpha$ and the relative error-bars as a function of different instrumental noises. Inspecting the results, we can highlight that, without considering center-to-limb CB, granulation and oscillations, $\alpha$ is always measurable for instrumental noise $\le 2$~m~s$^{-1}$. When $\alpha = 0.3$ and the white noise is $3$~m~s$^{-1}$, the error-bar is already large if compared to the input value, rendering the detection less constrained.

Injecting the center-to-limb CB, which is on a level of $\sim 1.75$~m~s$^{-1}$, underestimates the recovered $\alpha$ as the instrumental precision is $\le 1$~m~s$^{-1}$. Nonetheless, in general the results are compatible with the input values by $1-2 \sigma$ for the three parameters explored, $\alpha$, $\lambda$ and $v_* \sin i_*$. When the instrumental noise is $2-3$~m~s$^{-1}$ the retrieved $\alpha$ is not any more affected by CB. A low center-to-limb CB is justified by the fact that the star is a fast rotator and the effect of CB is diluted. This result is in line with the findings of \citet{Cegla16}, who mentioned that the CB dominates the residuals only for slow rotators. 

The injection of $\beta$Hyi granulation causes similar results to those we obtain for CB, since they are still on a level of $1.75$~m~s$^{-1}$, while the addition of the oscillations to the granulation introduces a red noise of $2.95$~m~s$^{-1}$ and increases the minimum $\chi^2$ with respect to the cases only with granulation. This impacts the accuracy and the uncertainty on the fit (see Tables \ref{resultsRMwithDRCBGRANOSC0.3} and \ref{resultsRMwithDRCBGRANOSC0.6}). In particular, it enlarges the error-bar on $\alpha$ and changes the best-fit results for the projected spin-orbit angle. Nonetheless, the retrieved alpha is closer to the injected one. To understand why accounting for just granulation affects more the results than having both granulation and oscillations, we inspected the mock RMs for the two cases. We observed that the granulation signal shows a lower frequency variation when compared to the oscillations. As a result, this changes the shape of the data, favouring lower values of $\alpha$. The additional signal from oscillations compensates for this effect, enlarging the uncertainty on $\alpha$. 

Finally, for all the performed tests, as the instrumental noise reaches $5$~m~s$^{-1}$ and $7$~m~s$^{-1}$, the error on $\alpha$ becomes large and, especially for $7$~m~s$^{-1}$, the detection of the differential rotation is no longer significant because the error-bar is equivalent to more than 50\% of the best fit $\alpha$. The error-bars do not significantly change when we add the stellar noises, indicating that white noise is the most determining factor for constraining the level of stellar differential rotation. The results at $5$ and $7$~m~s$^{-1}$ are in line with the test we performed on the WASP-7b observations, where the instrumental noise was $5.6$~m~s$^{-1}$. 

For comparison, we performed another test considering the resolution of PFS. We modelled the WASP-7b RM at the same phases as the real observations by \cite{Albrecht12}. We included the effect of differential rotation in the mock data with $\alpha = 0.45$, according to the best estimate of \cite{Albrecht12}. We also added a white noise with standard deviation equal to the average uncertainties in the real observations (5.6~m~s$^{-1}$). Then, we fitted this simulation and we obtained $\lambda = 84.4^{\circ} \pm 1.6^{\circ}$, $v_* \sin i_* = 13.5 \pm 2.0$~km~s$^{-1}$ and $\alpha = 0.47 \pm 0.45$. Comparing the error-bars of this test to those for the ESPRESSO case with $5$ and $7$~m~s$^{-1}$ and those for the test on real data, we observe a similar behaviour. The different resolution does not significantly affect the uncertainties.

We can conclude that $\alpha$ can be constrained if the white noise is $2$~m~s$^{-1}$ and and the stellar noise sources are lower than the instrumental noise. With 3~m~s$^{-1}$, $\alpha$ is detectable only if it is higher than $0.3$, but when the instrumental precision is 5~m~s$^{-1}$, the stellar differential rotation can no longer be constrained and a rigid body rotation cannot be ruled out.

To visualize an example of the fits so far performed, we report in Figure~\ref{Testgranosc} a plot of the simulation which included the differential rotation with $\alpha = 0.6$ and $\beta$ Hyi-like granulation and oscillations, compared to its best fit model. We see that considering these noises sources, the residuals are around 8~m~s$^{-1}$, lower than the previous observations of WASP-7b, where the residuals are 25~m~s$^{-1}$.

\subsection{Varying granulation and oscillations}
\label{varyinggranosc}
\begin{figure*}
\centering
\includegraphics[width=8.7cm]{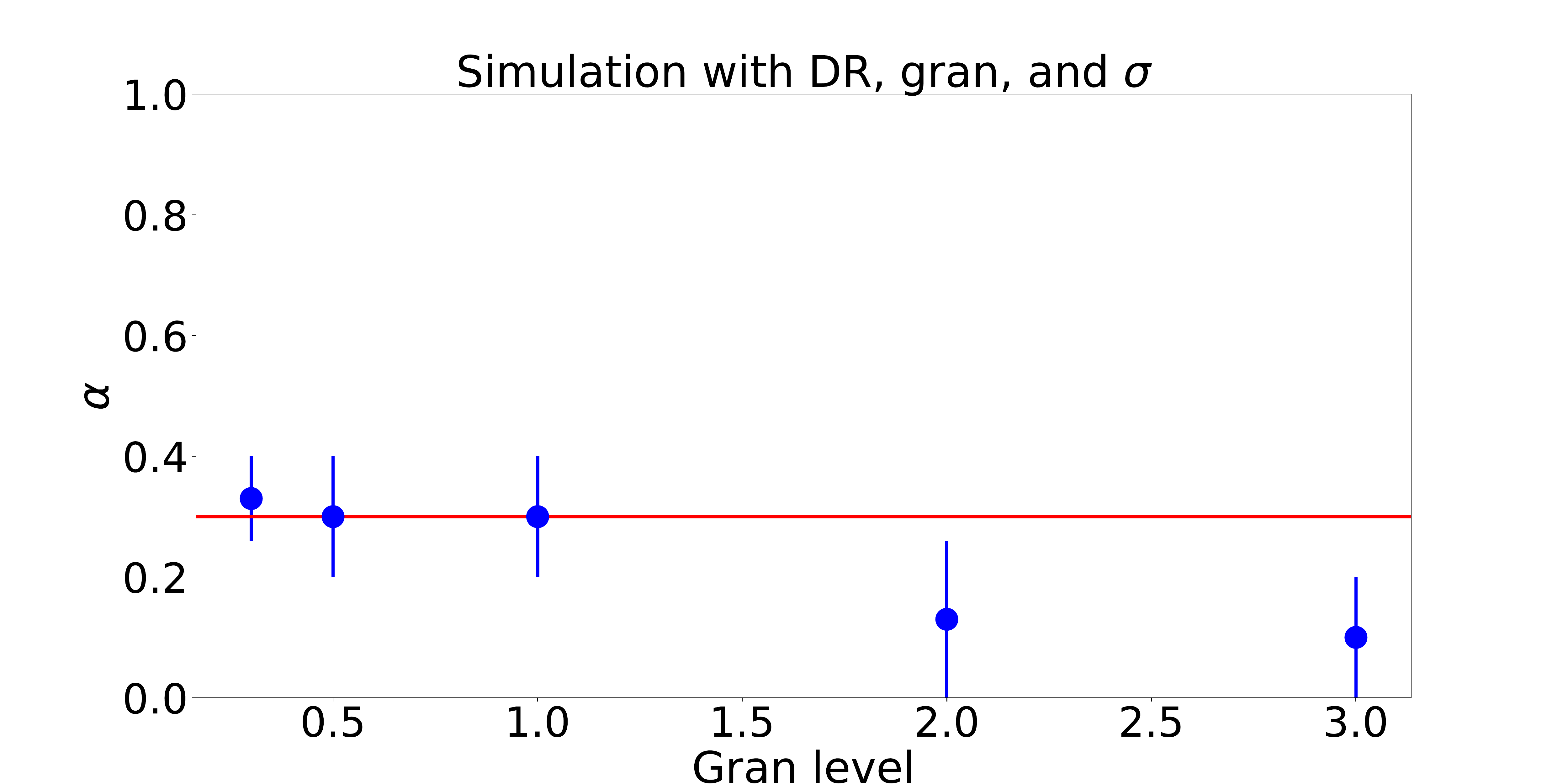}
\includegraphics[width=8.7cm]{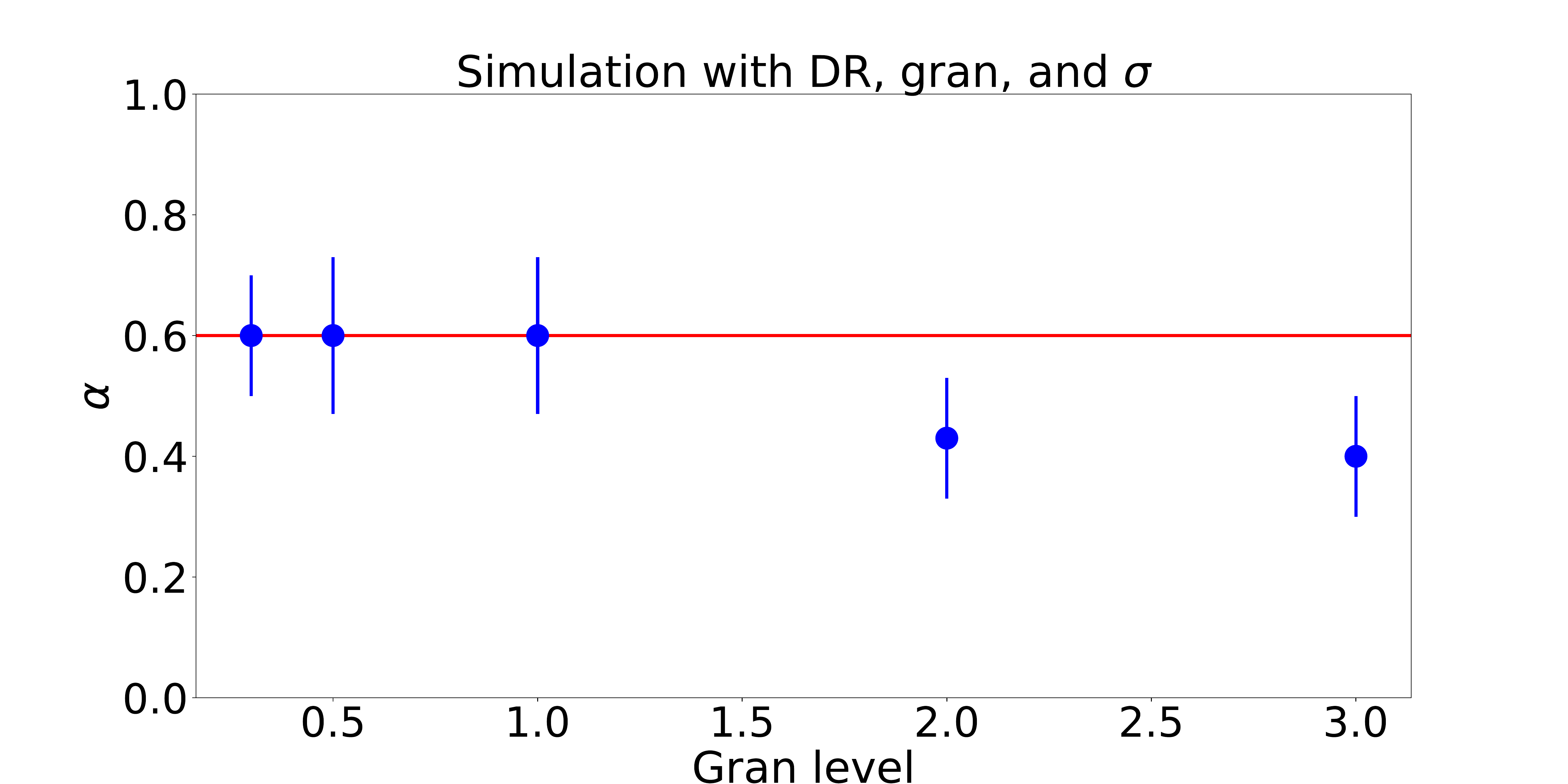}
\includegraphics[width=8.7cm]{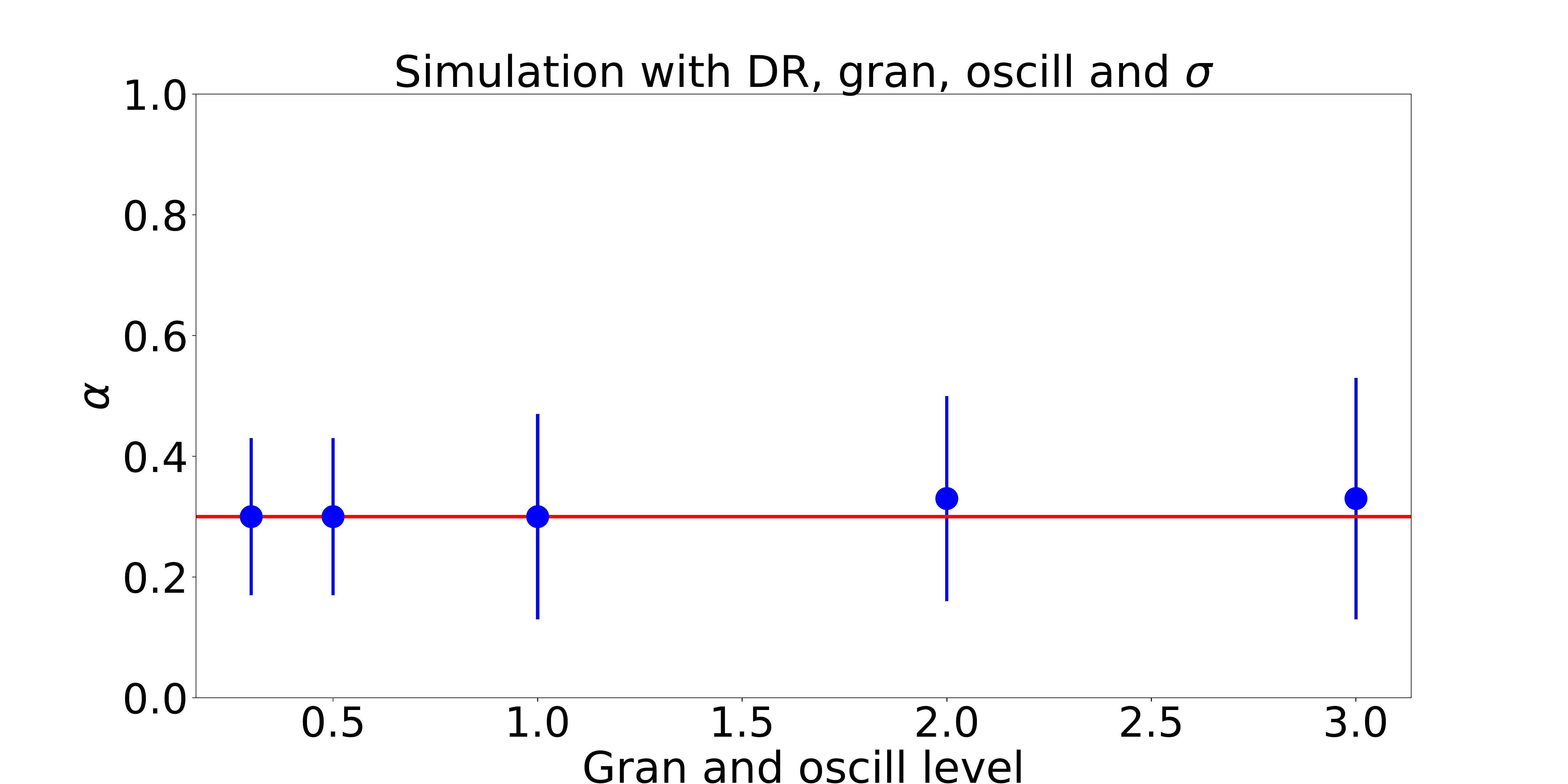}
\includegraphics[width=8.7cm]{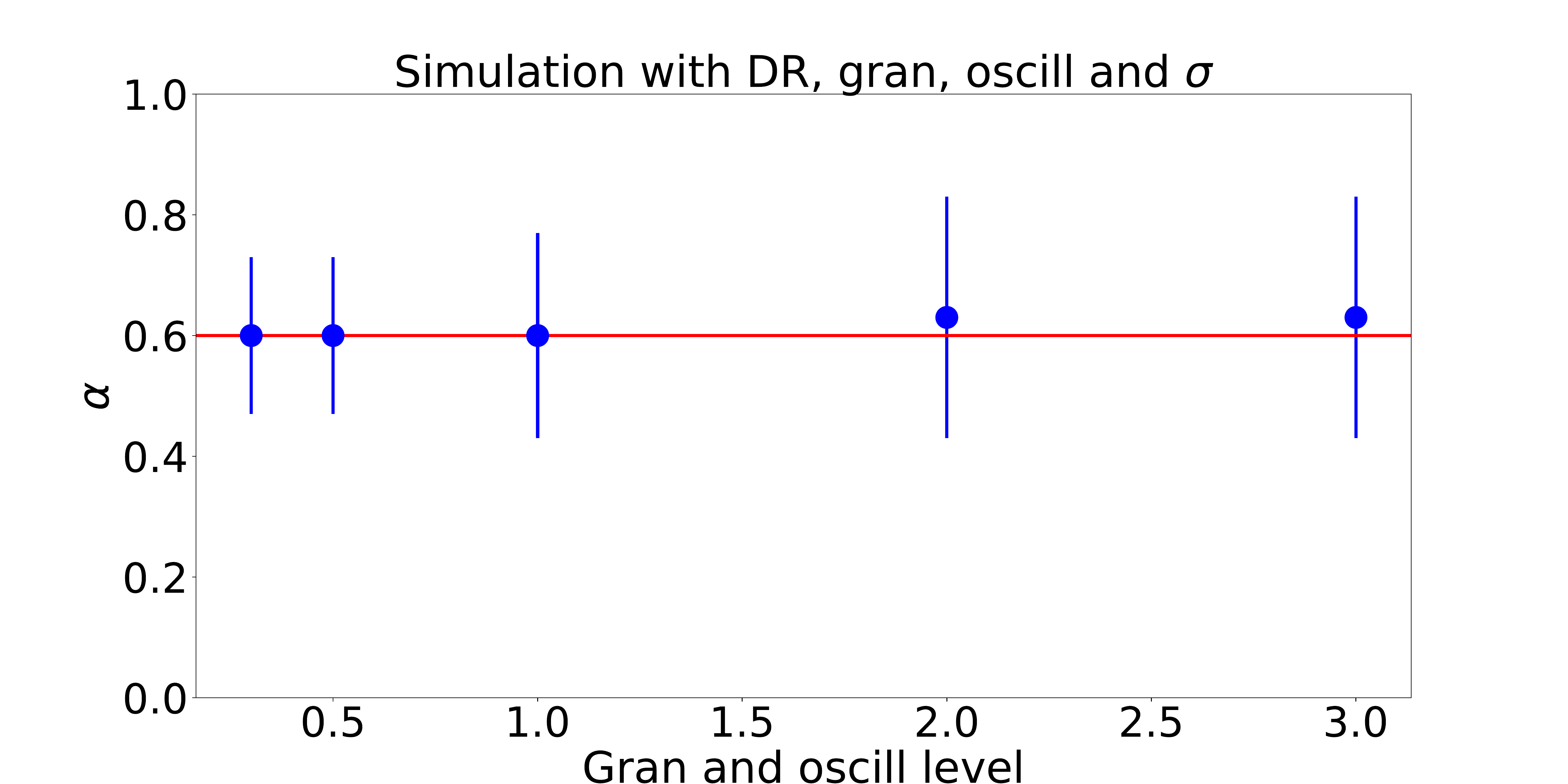}
\caption{Best fit $\alpha$ and relative error-bars as a function of the instrumental noise. In the first row, results for simulations of RM which included differential rotation (DR), granulation (gran) and instrumental noise ($\sigma$), in the second row for simulations including the oscillations (oscill) too. On the left side, plots relative to $\alpha = 0.3$, on the right side those for $\alpha = 0.6$. The fit accounts only for the differential rotation in the model.}
\label{Resultsgranlevel}
\end{figure*}

 $\beta$ Hyi is a G9 star and its granulation and oscillation-induced RV variability are predicted to be lower than those of WASP-7, because the frequencies of these phenomena depend on the spectral type and are higher for F-type stars \citep[around 60\% higher,][]{Dravins90}. In this section, we aim at exploring at what level the granulation and oscillations start to affect the detection of $\alpha$ if the instrumental noise is $2$~m~s$^{-1}$. As a first test, to account for different levels of granulation, we varied its amplitude, multiplying the simulated RVs described in Section~\ref{GRANOSCILL} by an amplification factor chosen among $0.33, 0.5, 1, 2$ and $3$. This corresponds to create an extra red noise on a level of $0.58, 0.875, 1.75, 3.5$ and $5.25$~m~s$^{-1}$ in the case of granulation, and of $0.98, 1.475, 2.95, 5.9$ and $8.85$~m~s$^{-1}$.  We produced RM simulations including instrumental noise of $2$~m~s$^{-1}$ and differential rotation for both $\alpha = 0.3$ and $0.6$. We then added the granulation and fitted the seven mock RMs we obtained. The results are in the left sides of Tables \ref{resultsRMwithDRvariousgran0.3} and \ref{resultsRMwithDRvariousgran0.6}. In the first row of Figure \ref{Resultsgranlevel}, we also show the dependence of the error-bar of the best fit $\alpha$ as a function of the amplification factor. As far as the amplification factor is lower than or equal to one, the granulation remains below the instrumental noise. The geometry of the system is slightly affected, with a maximum $\lambda$ variation of $1.6^{\circ}$ with respect to the input value. When the granulation is larger than the instrumental noise, all of the free parameters tend to decrease and the $\alpha$ significantly diverges from the starting value.

We repeated these tests for simulations including both granulation and oscillations together. The results of the fits are in the right sides of Tables \ref{resultsRMwithDRvariousgran0.3} and \ref{resultsRMwithDRvariousgran0.6}. In the second row of Figure \ref{Resultsgranlevel}, we show the dependence on the error-bar of the best fit $\alpha$ as a function of the amplification factor. As we noticed in the previous section, the addition of oscillations seems to compensate the changes induced by granulation. Although the error-bars on $\alpha$ are almost doubled and we registered a best $\chi^2$ higher than in the case only with granulation. The reasons for which adding the oscillations no longer affects the accuracy of $\alpha$ are similar to those described in section~\ref{varying_instrumental_noise}.

We can conclude that the granulation alone starts to affect as it becomes higher than that estimated for $\beta$ Hyi by \cite{Dumusque11}, while including both $\beta$ Hyi-like granulation and oscillations affects the precision with which we recover the differential rotation. Note that when we multiply by 3 the $\beta$ Hyi granulation and oscillations, the additional stellar noise is 8.85~m~s$^{-1}$, close to the extra RMS of 9.8~m~s$^{-1}$ we estimated fitting WASP-7b observations. 

\subsection{Varying the exposure time}
\label{exposure_time}
 All the previous simulations show that with an instrumental precision of $2$~m~s$^{-1}$ we have the opportunity to remove the white noise as a significant constrain in detecting the stellar differential rotation. Still, we mentioned that with 6~min. of exposure, the ESPRESSO predicted precision for WASP-7 is $3$~m~s$^{-1}$, due to its fast rotational velocity. One way to decrease the instrumental noise is observing for a longer exposure time. To examine how this could affect the final fit, we performed two tests, with an exposure time of 10 and 15~min, respectively. For the two cases, we estimated an ESPRESSO instrumental precision of 2.78~m~s$^{-1}$ and 2.27~m~s$^{-1}$ respectively. We produced again mock curves for WASP-7b, accounting for these uncertainties and decreasing the number of data points, to consider the longer exposure time. In the fits, we only accounted for the stellar differential rotation. We fitted the mock curves and obtained: 1) $\alpha = 0.59 \pm 0.22$, $v_*\sin i_* = 14.0 \pm 2.0$~km~s$^{-1}$ and $\lambda = 86.0^{\circ} \pm 1.6^{\circ}$, 2) $\alpha = 0.50 \pm 0.09$, $v_*\sin i_* = 13.5 \pm 0.5$~km~s$^{-1}$ and $\lambda = 86.0^{\circ} \pm 2.4^{\circ}$. For both of the cases, measuring the differential rotation is possible. When the exposure time is 10~min., $\alpha$ has a larger error-bar, which makes the determination of the stellar differential rotation less precise, even though more accurate. With 15 min of exposure, $\alpha$ is not any more accurate, however still compatible with the input value within $2\sigma$. 
 
 The choice to account for longer exposure times is justified by the attempt of overcoming the presence of granulation and oscillations. This is not entirely true. For example \cite{Dumusque11} say that for a G star 10~min. of integration are required to lower the oscillation to 50~cm~s$^{-1}$ and at least $30$~min to reduce the significance of granulation to the same level. F stars have stronger granulation and oscillations than G stars \citep{Dravins90}, thus integrating over 10 or 15~min is not enough to get rid of these phenomena. As mentioned in \cite{Chaplin}, for an F5V star as WASP-7 102 min. of exposure are required to reduce the oscillation to 0.1~m~s$^{-1}$. In the previous sections, we also find that as the granulation and oscillations contribution is higher than those estimated for $\beta$ Hyi, these phenomena are no longer negligible with a precision of $2$~m~s$^{-1}$, because they inject an extra RMS, and affect the estimation of $\lambda$ and $\alpha$. For all these reasons, when analyzing observations of WASP-7b, even with longer exposure times, ignoring these phenomena will determine systematic biases in the estimation of $\alpha$.

\subsection{The effect of convective broadening}

Another consequence of convection is a broadening of the CCF FWHM, which amounts to $\ge6$~km~s$^{-1}$ for F type stars \citep{Doyle}. This phenomenon is referred to as macro-turbulence. To test whether underestimating or overestimating it may affect the measurement of differential rotation via the RM effect, we produced simulations of the RM signal for WASP-7b. In the mock RMs we accounted for the stellar differential rotation, fixing $\alpha = 0.3$ and $0.6$, and for the instrumental noise, 2~m~s$^{-1}$. We varied the FWHM of the injected CCF, assigning values of $6.0$, $6.8$, $7.2$, $7.6$ and $8.0$~km~s$^{-1}$ and fitted the resulting mock data, assuming a FWHM $ = 6.4$~km~s$^{-1}$ (considering the instrumental resolution and the predicted macro-turbulence for WASP-7). The results are in Table \ref{resultsRMwithDRandFWHM0.3}, which show that the change of the local FWHM slightly affects the parameters, although still the best fit values are compatible with the injected ones within $1 \sigma$. As the FWHM increases, the best fit $v_* \sin i_*$ decreases. This result is reasonable because an enlargement of the CCF is compensated by a lower depth of the CCF itself and in good compatibility with the findings of \cite{Doyle}, who showed that for fast rotators not accounting properly for the macro-turbulence slightly affects the best-fit, injecting uncertainties of $0.5$~km~s$^{-1}$ on $v_*\sin i_*$. 

\subsection{Limb darkening effect}
Injecting improper limb darkening coefficients might as well affect the measurement of $\alpha$. To explore to which extend this could affect the estimated $\alpha$, we produced simulations of the WASP-7b RM signal accounting only for the instrumental noise of $2$~m~s$^{-1}$ and the differential rotation ($\alpha = 0.3$ and $0.6$). We modified the limb darkening coefficients from assigning parameters selected from the table by \citet{Claret} for F stars. The coefficients were selected based on our original input values, $u_1 = 0.2$ and $u_2 = 0.3$, and considering typical uncertainties were in line with the typical uncertainties of $0.05-0.1$ as seen in literature \citep[e.g.][]{Albrecht12, Addison18}. In particular, we chose: $u_1 = 0.25$ and $u_2 = 0.25$, $u_1 = 0.15$ and $u_2 = 0.35$, and also $u_1 = 0.25$ and $u_2 = 0.35$. We then fitted each simulation, using as limb darkening parameters in the fitting model $u_1 = 0.2$ and $u_2 = 0.3$. The results are shown in Table \ref{resultsRMwithDRandLIMB0.3}.

All the results we obtained show compatibility with the injected parameters within 1$\sigma$. As long as the limb darkening coefficients are changed so that if one increases the other decreases, then the recovered $\alpha$ is the same as the injected value. In the last test, we increased both $u_1$ and $u_2$; the recovered $\alpha$ was different than the input one, even if still compatible with the input value within $1\sigma$. The fact that only $\alpha$ is affected is reasonable. Since the planet has a high projected spin-orbit angle, during the transit it occults nearly the same longitude. In this sense, the greatest changes we observe in $v_* \sin i_*$ occur on the transit ingress and egress, because here the planet is closer to the stellar pole, which rotate slower due to the differential rotation. For this reason varying the limb darkening coefficient mainly affects the retrieved $\alpha$. This explanation is valid only for the case in which $\lambda$ is close to $90^{\circ}$ \citep[see e.g.][for a different geometrical configuration, with an almost aligned planet]{Cegla16}. 

We can conclude that within reasonable uncertainties, the limb darkening does not significantly affect the determination of the stellar differential rotation - at least for highly inclined systems like WASP-7. Moreover, the limb darkening coefficients are usually well constrained from transit observations, usually available when exploring the RM signal generated by an exoplanet.

\subsection{Spots}
The determination of differential rotation could be plagued by occulted spots or other magnetic activity related features (e.g. faculae). To explore this aspect, we isolated the residuals of the fit of WASP-7b RM with $\alpha = 0.6$ and an instrumental noise of $2$~m~s$^{-1}$ and tried to reproduce it with spots and plages crossing the transit chord. The only condition wherein the stellar activity features might resemble the residuals was by having at least two spots on the transit ingress and egress and several plages along the transit chord, which seems unlikely, also given that the plage are more often visible on the limbs. Nonetheless, if during the transit the planet crosses a spot on the stellar disc, the RM signal shows a bump, which cannot be reproduced with differential rotation and needs to be accounted for to get the correct system geometry.

Non-occulted spots change the shape and depth of the RM and their effect on the estimate of system geometry was already explored in \cite{Oshagh18}, to which we refer for more details. 

\section{Discussion and conclusions}
\label{discussion}
In this work, we explored the possible reasons behind the hurdles in detecting the stellar differential rotation through RM observations. On top of this, we analyzed the improvements we could gain by adopting more precise and stable spectrographs, on larger aperture telescopes (e.g. ESPRESSO).

We first fit the observed RM of WASP-7 with a model which accounted for the stellar differential rotation and the instrumental noise reported in \citet{Albrecht12}. The result gave an $\alpha$ with a large errorbar, and a system geometry in full agreement with the result reported in \citet{Albrecht12b}. Moreover, we performed a series of tests, in which we modelled the WASP-7 RM signal with different levels of instrumental noise, varying $\alpha$ and eventually accounting for certain sources of stellar noise (granulation, oscillation, center-to-limb CB, limb darkening and convective broadening/macro-turbulence). We obtained the following results:
\begin{itemize}
\item in the absence of any stellar noise source, if the instrumental noise is $5$~m~s$^{-1}$ or more, $\alpha$ is no longer detectable. The addition of stellar noise affects the results by enlarging the error-bars. Moreover, for a white noise of $7$~m~s$^{-1}$, the injection of stellar noise caused an underestimation of the retrieved $\alpha$ by at least $0.1$.
\item for an instrumental noise of 2-3~m~s$^{-1}$, a solar-like center-to-limb CB and a $\beta$Hyi-like granulation do not bias the detection of the stellar differential rotation. The final results are compatible with those retrieved in the absence of stellar noise within $1\sigma$.
\item if the instrumental noise is lower than 2~m~s$^{-1}$, adding the center-to-limb CB or granulation leads to an underestimated $\alpha$. Adding the oscillations to the simulations with granulation, the value of alpha is closer to the injected value, though with higher uncertainties. A closer inspections of the mock RMs shows that this happens because of a lower frequency variation in the granulation component, if compared to the case with both granulation and oscillations. 
\item for a white noise of 2~m~s$^{-1}$, if granulation and oscillations together are twice or three times the $\beta$Hyi level, they cause an underestimation of the projected spin-orbit angle by at least $2.5^{\circ}$ and enlargement of the error-bar on $\alpha$. Hence, these noise sources need to be accounted for, when modelling data with 2~m~s$^{-1}$ or better precision. The granulation alone causes an underestimate of both $\alpha$ and $v_*\sin i_*$. The reasons for these different results between the simulations only with granulation and those with both granulation and oscillations are similar to those explained in the previous point.
\item the convective broadening slightly affects the $v_*\sin{i_*}$ when the instrumental noise is 2~m~s$^{-1}$. In detail, it decreases the retrieved projected velocity as the FWHM of the CCF becomes larger
\item varying the limb darkening parameters by 0.05 each overestimates $\alpha$ by maximum $0.1$ when the instrumental noise is 2~m~s$^{-1}$.
\item for a WASP-7 like system, in the absence of any stellar noise, at 2~m~s$^{-1}$ the minimum detectable $\alpha$ is 0.2. If the instrumental noise increases to 3~m~s$^{-1}$ and stellar noise is included in the mock data, the lowest detectable $\alpha$ increases to more than $0.3$.
\end{itemize}

These results suggest that, even with a precise knowledge of the stellar noise of WASP-7, the PFS instrumental noise of $5.6$~m~s$^{-1}$ by itself imposes a limit to the detection of the stellar differential rotation. On top of this, as we showed in Section~\ref{wasp-7b}, a fit to WASP-7b PFS observations showed an average residual level of 11~m~s$^{-1}$. A difference in quadrature between this and the reported instrumental noise ($5.6$~m~s$^{-1}$) gave an additional RMS of 9.95~m~s$^{-1}$. This high level is hard to justify with any of the stellar noise sources we explored in this paper. If they do happen to have a stellar origin, they will be present as well in the data retrieved with a more stable spectrograph and therefore they could still bias the detection of $\alpha$. Nonetheless, a stable spectrograph could allow us to more easily disentangle stellar and instrumental variability and, therefore, we could account for both in the fitting model. 

A close inspection of the PFS residuals shows a similar behaviour between the in and out of transit signal, which allows us to exclude a few possible stellar noise sources not inspected in the present analysis:
\begin{itemize}
\item a higher center-to-limb CB than the solar-like one. The similarity between the in and out of transit residuals, excludes the possibility that a stronger CB might be the source of a significant fraction of the additional RMS.
\item some occulted and non-occulted spots might change the shape and depth of the transit, and bias the measured geometry \citep{Oshagh18}. Nonetheless, the rotational velocity of the star is 5 days. Thus spots could have not moved significantly over the course of a transit, nor evolved much.
\end{itemize}
This leaves three possibilities, which, combined, might generate the extra noise:
\begin{itemize}
\item a higher granulation than the $\beta$Hyi case. For F stars the granulation is predicted to be $60$\% higher than in the case of G stars \citep{Dravins90}. Considering both granulation and residual oscillations, the observations of WASP-7 could have at least 5~m~s$^{-1}$ of extra noise. Still, this value is far from covering the entire RMS of 9.95~m~s$^{-1}$, unless WASP-7 has a very high stellar granulations, which so far has never been predicted to exist and has never been detected. 
\item unaccounted instrumental noise, maybe connected to intrinsic instrumental instabilities.
\item weather condition changes (no information about SNR and seeing variation during observation is reported in \citet{Albrecht12} and they could lead to increase the RMS residuals).
\end{itemize}

We conclude that while the white noise represents a bias in detecting stellar differential rotation, we still have no definitive answer about the source of additional residuals in WASP-7b observations. Because of its stable and precise RVs, the new spectrograph ESPRESSO offers an important opportunity to understand if these residuals still remain in the observations and eventually understand their origins. Moreover, the large aperture of the Very Large Telescope (VLT) will allow for an increased sampling in the RM observations, which will aid in constraining potential stellar models. In this paper, we showed that for WASP-7 an instrumental error of 2~m~s$^{-1}$ or lower is necessary for detecting the stellar differential rotation. This level of precision can be reached with ESPRESSO in the 1UT mode for a star rotating slower than 10~km~s$^{-1}$ with a 6~min exposure time, however for stars rotating faster than 10~km~s$^{-1}$ a longer exposure time is necessary to decrease the instrumental noise \citep{Bouchy}. We showed that, for WASP-7, an exposure of 15~min should allow one to decrease the white noise to $2$~m~s$^{-1}$. However, in Section~\ref{exposure_time} we highlight that, with this exposure time, the recovered $\alpha$ is underestimated by $0.1$. ESPRESSO in the 4UT mode would be able to reach a precision of $2$~m~s$^{-1}$ in much less time, although it is unlikely that this mode will be used for RM observations alone. An alternative and more practical method could be to observe more transits and fold them one on top of the other. This would give more data points and decrease the uncertainty on $\alpha$. The best approach would be to apply this method on magnetically quiet stars. In the case of active stars, observing more transits would allow us to perform predictions on the spot contribution and account for them in the fit.

Once the instrumental noise is as low as 2~m~s$^{-1}$, and, if we can mitigate the possible noise sources encountered with the PFS observations, we may be able to measure the stellar differential rotation for WASP-7 and all stars with planetary system geometries similar to WASP-7b (from the TEPCAT catalog \cite{Southworth11} we cite CoRoT-1b, WASP-79b, WASP-100b, WASP-109b and HAT-P-32b). WASP-7 is an F star, thus on average it should have $\alpha = 0.1-0.2$. On one hand, if ESPRESSO-level RM observations in the future are unable to constrain the stellar differential rotation, we can conclude that this star has a maximum $\alpha = 0.2$. On the other hand, if such observations revealed an $\alpha = 0.3$ or higher, we could break the degeneracy between the stellar inclination and the stellar rotation, as anticipated in \cite{Hirano} and attempted in \cite{Cegla}. 

\section*{acknowledgements}
This work was supported by FCT/MCTES through national funds (PIDDAC) by the grant UID/FIS/04434/2019. This work was also supported by FCT - Funda\c{c}\~ao para a Ci\^encia e a Tecnologia through national funds (PTDC/FIS-AST/28953/2017, PTDC/FIS-AST/32113/2017) and by FEDER - Fundo Europeu de Desenvolvimento Regional through COMPETE2020 - Programa Operacional Competitividade e Internacionaliza\c{c}\~ao (POCI-01-0145-FEDER-028953, POCI-01-0145-FEDER-03211). L.M.S. also acknowledges support by the fellowship SFRH/BD/120518/2016 funded by FCT (Portugal) and POPH/FSE (EC). M.O. acknowledges research funding from the Deutsche Forschungsgemeinschft (DFG, German Research Foundation)-OS 508/1-1. M.O also acknowledges the support of COST Action TD1308 through STSM grant with reference Number:STSM-D1308-050217-081659. H.M.C. acknowledges financial support from the National Centre for Competence in Research (NCCR) PlanetS supported by the Swiss National Science Foundation (SNSF). L. M. S., M.O, J.P.F, and B. A also acknowledge the support of the FCT/DAAD bilateral grant 2019 (DAAD ID: 57453096). N.C.S. and S.C.C.B. also acknowledge support from FCT through Investigador FCT contracts n$^{\circ}$s IF/00169/2012/CP0150/CT0002 and IF/01312/2014/CP1215/CT0004. A.B. also acknowledges support by the fellowship PD/BD/135226/2017 funded by FCT (Portugal) and POPH/FSE (EC).

\bibliographystyle{mnras}
\bibliography{bibliography}

%
\appendix
\onecolumn
\section{The effect of differential rotation on the RM signal}
\label{Appendix1}
\begin{figure*}
\centering        \includegraphics[width=13cm]{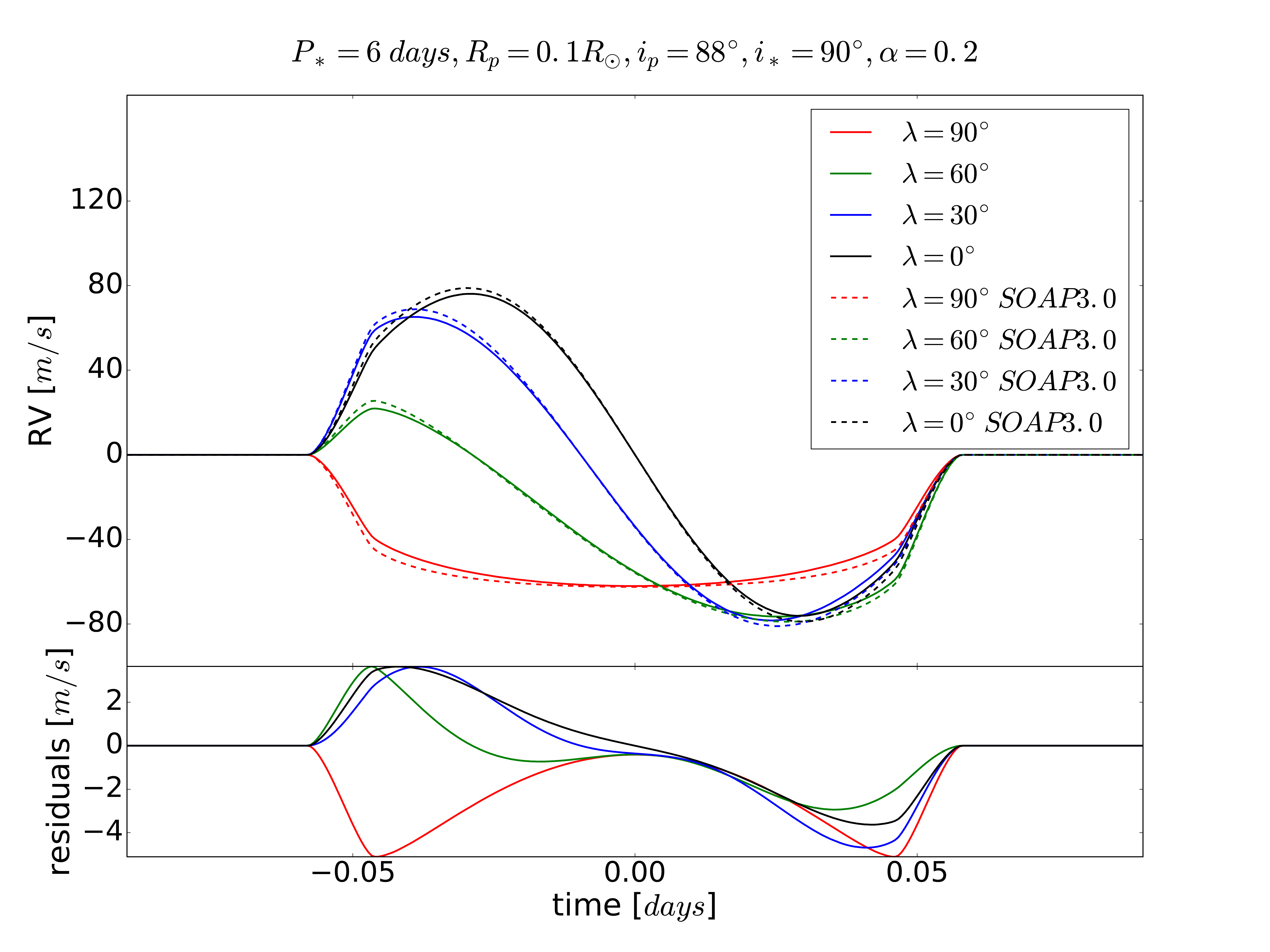}
    \caption{RM simulations for different values of $\lambda$, projected spin-orbit angle, $90^{\circ}$, $60^{\circ}$, $30^{\circ}$ and $0^{\circ}$. The dashed lines represent the same simulations, produced with the old version of \textit{SOAP3.0}, which did not implement the stellar differential rotation. In the bottom frames, we show the residuals with respect to the rigid rotation case.}
\label{Results_1}
\end{figure*}

\begin{figure*}
\centering
\includegraphics[width=13cm]{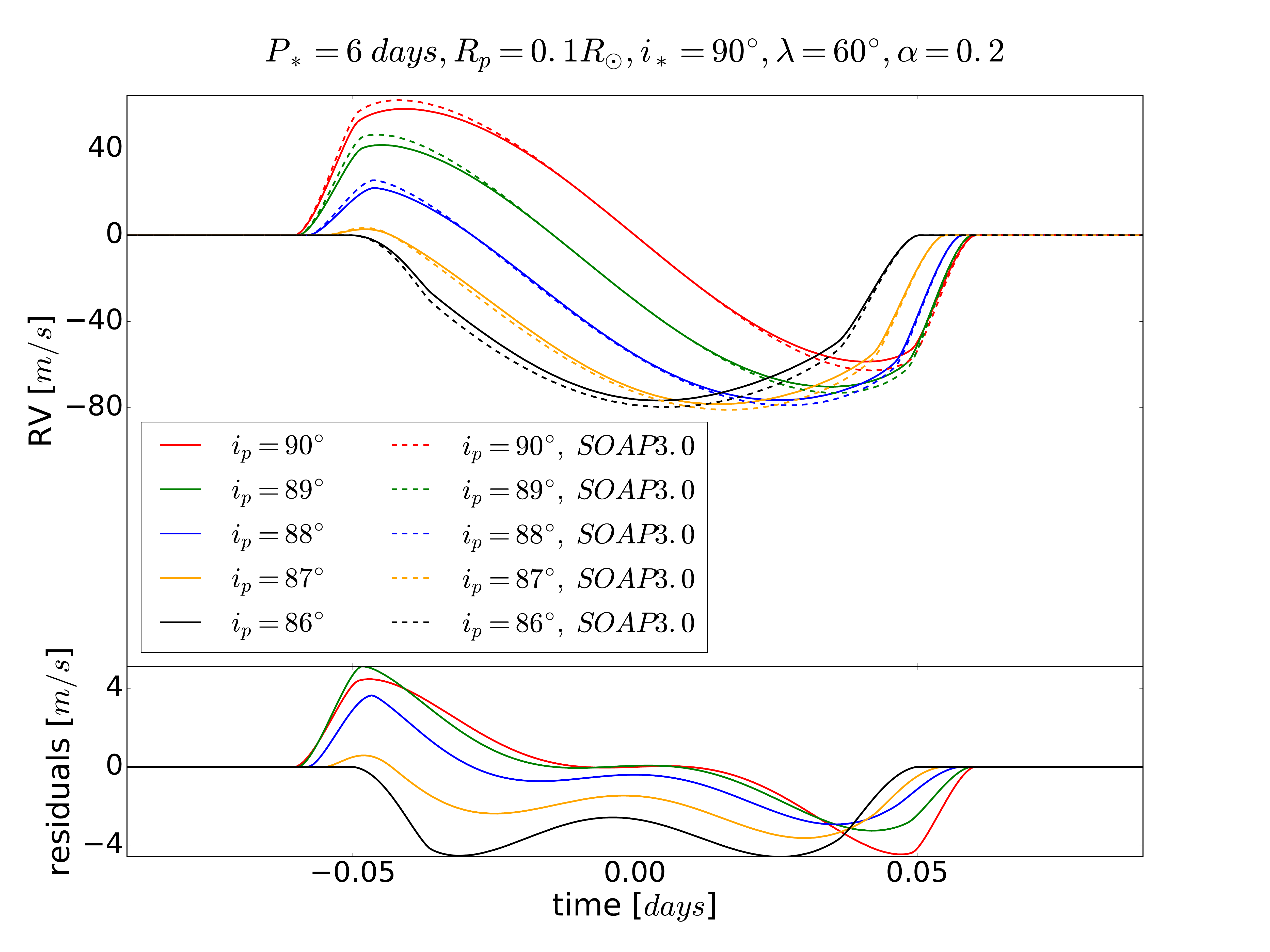}
\caption{RM simulations for different values of $i_P$, the planet orbital inclination, $90^{\circ}$, $89^{\circ}$, $88^{\circ}$, $87^{\circ}$ and $86^{\circ}$. The dashed lines represent the same simulations, produced with \textit{SOAP3.0}. The bottom part of each frame reports the residuals of the RM simulation with respect to the one produced with \textit{SOAP3.0}.}
\label{Results_2}
\end{figure*}

We performed a series of tests with the objective of determining how the differential rotation changes the RM signal once we vary the rotational pattern of the star and the planetary system geometry. We produced simulations with the following initial conditions: a stellar rotation of $P_* = 6$~days (to choose rapid rotators), a stellar inclination of $i_* = 90^{\circ}$ (i.e. the spin-axis is perpendicular to the line of sight), and a planet inclination of $i_P = 88^{\circ}$ (so the planet transits close to a stellar latitude of $45^{\circ}$). We fixed as well the planet radius to $R_p = 0.1 R_{\odot}$, to account for Jovian planets whose RM signal should be stronger than that by Earth- or Neptune-like planets. We finally fixed a projected spin-orbit angle of $\lambda = 60^{\circ}$, forcing the planet to mainly occult the red-shifted or blue-shifted side of the planet. In this way, the RM assumes an asymmetric shape. 

In Figure \ref{Results_1}, we show the results obtained by fixing $\alpha = 0.2$ and varying the projected spin-orbit angle. We found no significant difference between the cases with $\lambda = 60^{\circ}$, $30^{\circ}$ and $0^{\circ}$. However, for $\lambda = 90^{\circ}$, the RM signal is totally blue-shifted, meaning the planet only crosses the red-shifted side of the star. The residuals are stronger in the case of of $\lambda = 90^{\circ}$,with residuals that reach 4~m~s$^{-1}$. 

In Figure \ref{Results_2}, we report the simulations obtained with fixed $\alpha = 0.2$ and modified $i_P$, assigning values $90^{\circ}$, $89^{\circ}$, $88^{\circ}$ , $87^{\circ}$ and $86^{\circ}$. As $i_P$ decreases, the amplitude of the residuals tends to increase as well, because the planet occults an area of the stellar disk with a slower rotation than the equator. On the other hand, the length of the transit feature also decreases, because the transit happens on a shorter transit chord (the impact parameter $b$ increases). 

\begin{figure*}
\centering
\includegraphics[width=13cm]{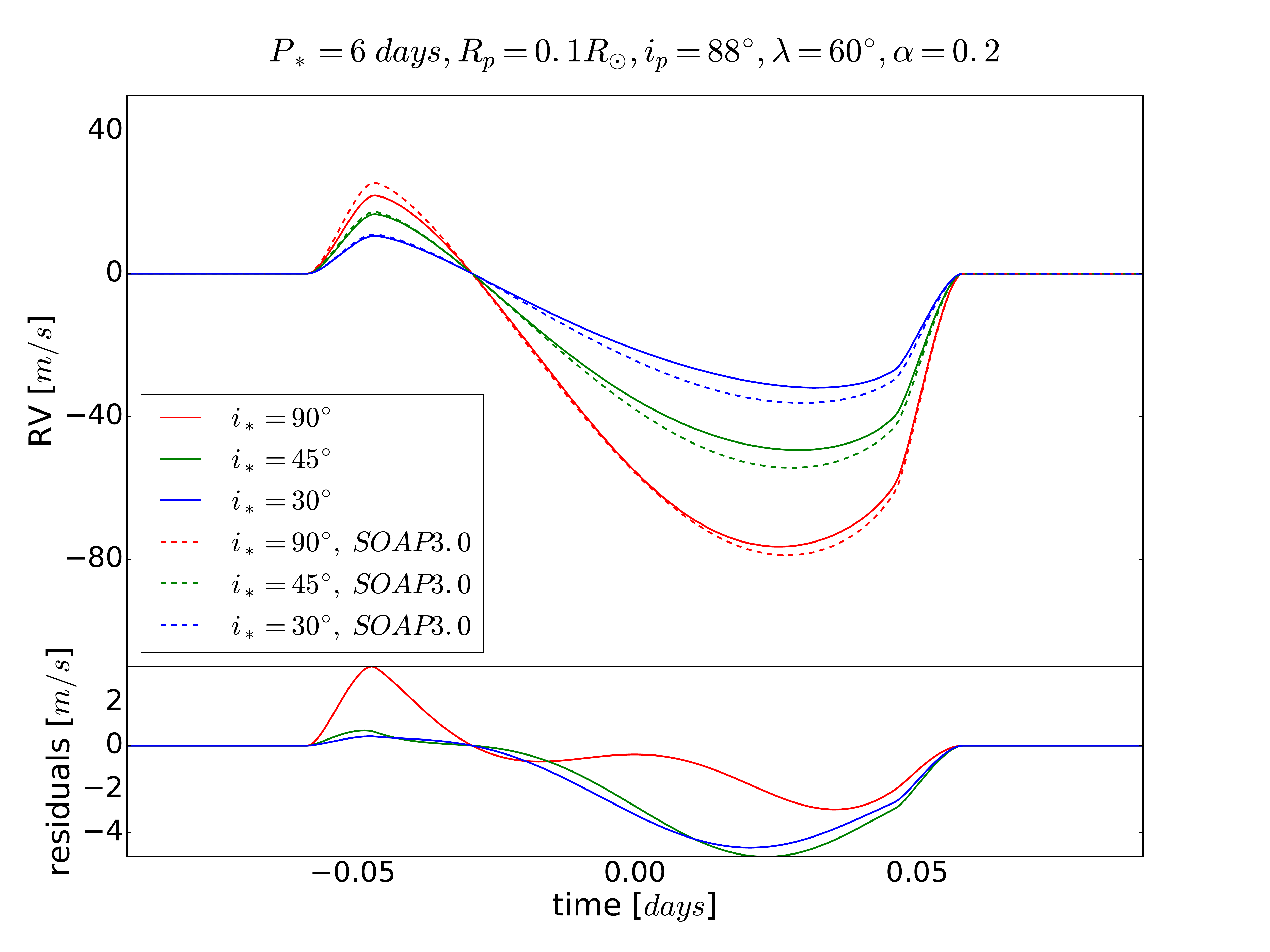}
\caption{RM simulations for different values of $i_*$, the inclination of the stellar rotational axis, $90^{\circ}$, $45^{\circ}$, $30^{\circ}$. The dashed lines represent the same simulations, produced with the old version of \textit{SOAP3.0} (without stellar differential rotation). The bottom frame reports the residuals of the RM simulation with respect to the one produced with \textit{SOAP3.0}.}
\label{Results_3}
\end{figure*}

\begin{figure*}
        \centering
        \includegraphics[width=13cm]{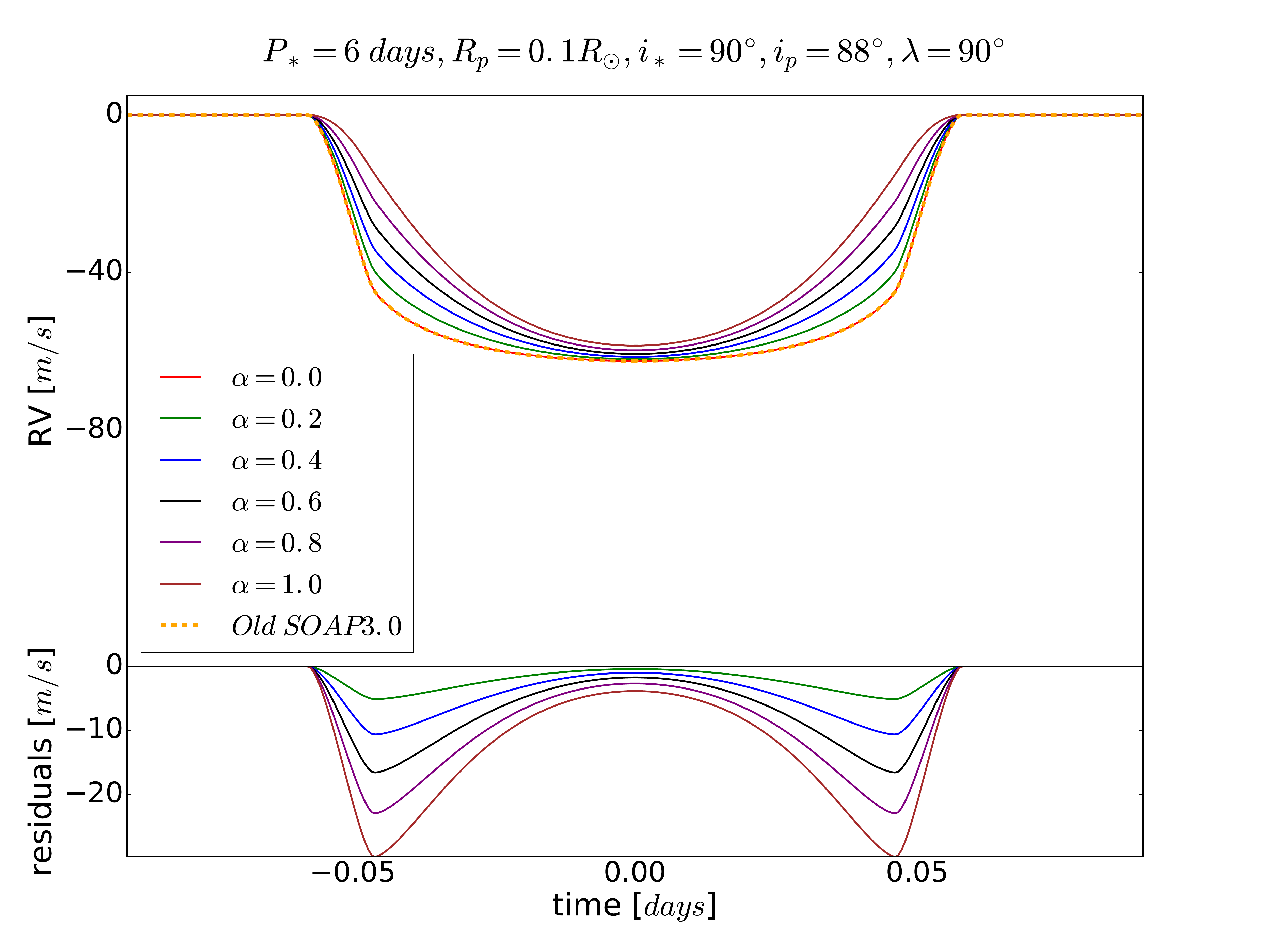}
        \includegraphics[width=13cm]{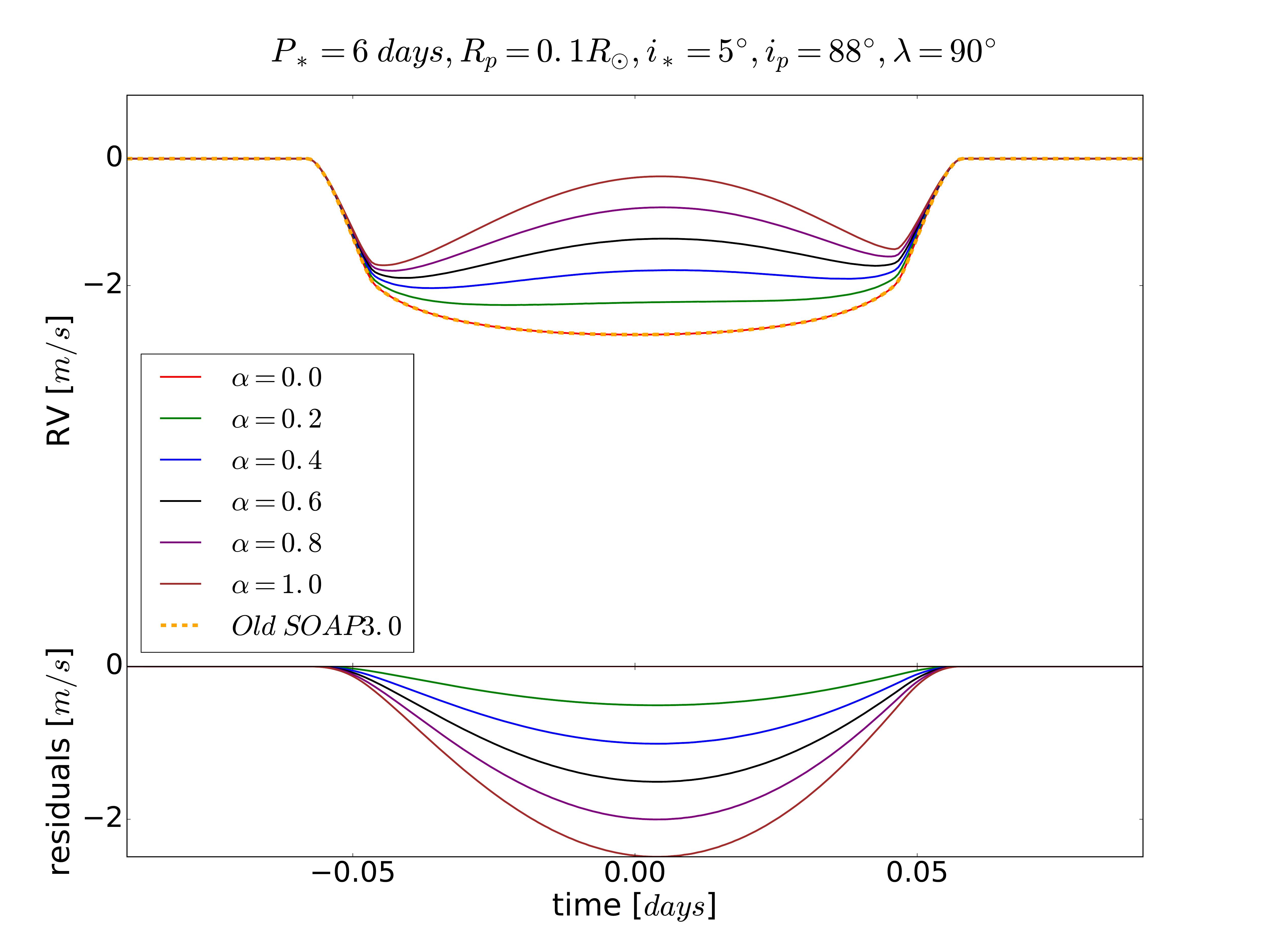}
\caption{RM  simulations  for  extreme  values  of $i_*$, $90^{\circ}$, in the top frame, and $5^{\circ}$, in the bottom one, to produce equator on and almost pole on  configurations, varying the differential rotation parameter $\alpha$. The dashed lines represent the same simulations, produced with the old version of \textit{SOAP3.0}. The bottom part of each frame reports the residuals of the RM simulations with respect to those without stellar differential rotation.}
\label{Results_4}
\end{figure*}

On top of this, we wanted to explore the effect of changing the inclination of the stellar rotational axis in the model. We first simulated the case with $P_* = 6$~days, $i_P = 88^{\circ}$, $R_P = 0.1R_{\odot}$ , $\lambda = 60^{\circ}$ and $\alpha =$ 0.2 and we varied the stellar inclination from $90^{\circ}$ to $45^{\circ}$, $30^{\circ}$ and $0^{\circ}$. The results are in Figure \ref{Results_3}. As the stellar inclination decreases, the planet transits slower rotating areas of the star and the RM residuals tend to increase, the effect of the stellar rotation is stronger. 

Thus, we decided to explore in detail how the residuals change if the stellar inclination is close to $0^{\circ}$ or $90^{\circ}$ and the projected spin-orbit angle is $\lambda = 90^{\circ}$ . In these conditions, the stellar axis and the planetary orbit are perpendicular. The plots in Figure \ref{Results_4} report the two cases with $i_* = 5^{\circ}$ in the top frame, and $90^{\circ}$ in the bottom one. The case with $i_* = 5^{\circ}$ is an almost pole-on configuration, in which the planet crosses areas of the stellar surface that rotate very slowly. The differential rotation causes the formation of a bump in the center of the RM. This bump happens in correspondence to the point of the transit where the planet passes closer to the pole of the star. As $\alpha$ increases, the bump becomes stronger. In the case, of $i_* = 90^{\circ}$, the residuals tend to increase on the stellar limbs, where the planet crosses areas of the star which rotate more slowly. No significant change can be observed in correspondence to the center of the transit, where the planet obscures an equatorial area. By analyzing these configurations, we can deduce that the differential rotation allows us to break the degeneracy between the stellar rotation at the equator and the inclination of the stellar rotational axis, because for different $i_*$ we obtain different residuals due to the geometry of the system. These results are in agreement with those by \citet{Hirano} and encourages a deeper analysis on the opportunity we have to analyze the rotational pattern of stars through the RM signal.

\section{Best fit values}
\label{Appendix2}
We report here the results we obtained for the different tests performed. In the left sides of Tables \ref{resultsRMwithDR0.3} and \ref{resultsRMwithDR0.6} we show the best fit values for the tests performed on simulations which included only differential rotation and instrumental noise. In the right sides of the same tables we report the results for the tests on simulations with both differential rotation and CB. The left side of Tables \ref{resultsRMwithDRvariousgran0.3} and \ref{resultsRMwithDRvariousgran0.6} lists the best fits of the tests on simulations of RM which include differential rotation, CB, different levels of granulation and instrumental noise. In the right side of the same tables we show the results for the case in which the oscillations where also included in the model. In the left side of Tables \ref{resultsRMwithDRCBGRANOSC0.3} and \ref{resultsRMwithDRCBGRANOSC0.6} we report the results obtained by fitting simulations of RM including stellar differential rotation, CB, granulation and different levels of instrumental noise, while in the right side the fitted simulations also included oscillations. 
Table \ref{resultsRMwithDRandFWHM0.3} lists the results for the tests with a varied FWHM, while in Table \ref{resultsRMwithDRandLIMB0.3} we report the results for the tests in which we varied the limb darkening coefficients.
\begin{table*}
\caption{Results of our fitting procedure for the WASP-7 RM simulations: 1) including instrumental noise and differential rotation (DR); 2) including instrumental noise, center-to-limb convective blue-shift (CB) and differential rotation (DR). The injected $\alpha$ was $0.3$ and the fit was performed accounting for differential rotation.}             
\label{resultsRMwithDR0.3}      
\centering                          
\begin{tabular}{ c c c c c c c c c }        
\hline\hline
Simulation & &  \multicolumn{3}{c}{DR} & & \multicolumn{3}{|c}{DR + CB } \\
\hline\hline
noise  && $v_*\sin i_*$ & $\lambda$ & $\alpha$ && $v_*\sin i_*$ & $\lambda$ & $\alpha$  \\  
\hline 
$0.50$~m~s$^{-1}$ && $14.0 \pm 0.1$~km~s$^{-1}$ & $86.0^{\circ} \pm 0.4^{\circ}$ & $0.30 \pm 0.03$ && $13.8 \pm 0.3$~km~s$^{-1}$ & $86.0^{\circ} \pm 0.8^{\circ}$ & $0.23 \pm 0.03$ \\
$1.00$~m~s$^{-1}$ && $14.0 \pm 0.3$~km~s$^{-1}$ & $86.0^{\circ} \pm 0.8^{\circ}$ & $0.30 \pm 0.03$ && $13.8 \pm 0.3$~km~s$^{-1}$ & $86.0^{\circ} \pm 0.8^{\circ}$ & $0.23 \pm 0.10$ \\
$2.00$~m~s$^{-1}$ && $14.0 \pm 0.3$~km~s$^{-1}$ & $85.2^{\circ} \pm 1.2^{\circ}$ & $0.30 \pm 0.10$ && $14.0 \pm 0.5$~km~s$^{-1}$ & $85.2^{\circ} \pm 1.2^{\circ}$ & $0.30 \pm 0.13$ \\
$3.00$~m~s$^{-1}$ && $14.0 \pm 1.0$~km~s$^{-1}$ & $84.2^{\circ} \pm 1.6^{\circ}$ & $0.27 \pm 0.13$ && $14.0 \pm 1.0$~km~s$^{-1}$ & $84.8^{\circ} \pm 1.6^{\circ}$ & $0.27 \pm 0.20$ \\
$5.00$~m~s$^{-1}$ && $14.0 \pm 1.5$~km~s$^{-1}$ & $84.8^{\circ} \pm 1.6^{\circ}$ & $0.30 \pm 0.23$ && $14.0 \pm 1.5$~km~s$^{-1}$ & $86.6^{\circ} \pm 2.4^{\circ}$ & $0.30 \pm 0.23$ \\
$7.00$~m~s$^{-1}$ && $14.0 \pm 1.5$~km~s$^{-1}$ & $86.0^{\circ} \pm 4.0^{\circ}$ & $0.40 \pm 0.24$ && $13.3 \pm 2.3$~km~s$^{-1}$ & $86.0^{\circ} \pm 3.2^{\circ}$ & $0.13 ^{+0.40}_{0.13}$ \\
\hline
                            
\end{tabular}
\end{table*}

\begin{table*}
\caption{Same as in Table \ref{resultsRMwithDR0.3} but for $\alpha = 0.6$.}             
\label{resultsRMwithDR0.6}      
\centering                          
\begin{tabular}{ c c c c c c c c c }        
\hline\hline
Simulation &&  \multicolumn{3}{c}{DR} && \multicolumn{3}{|c}{DR + CB} \\
\hline\hline 
noise  && $v_*\sin i_*$ & $\lambda$ & $\alpha$ && $v_*\sin i_*$ & $\lambda$ & $\alpha$  \\  
\hline 
$0.50$~m~s$^{-1}$ && $14.0 \pm 0.3$~km~s$^{-1}$ & $86.0^{\circ} \pm 0.4^{\circ}$ & $0.60 \pm 0.03$ && $13.8 \pm 0.3$~km~s$^{-1}$ & $86.0^{\circ} \pm 0.4^{\circ}$ & $0.53 \pm 0.03$ \\
$1.00$~m~s$^{-1}$ && $14.0 \pm 0.3$~km~s$^{-1}$ & $86.0^{\circ} \pm 0.8^{\circ}$ & $0.60 \pm 0.03$ && $13.8 \pm 0.3$~km~s$^{-1}$ & $86.0^{\circ} \pm 0.8^{\circ}$ & $0.53 \pm 0.03$ \\
$2.00$~m~s$^{-1}$ && $14.0 \pm 0.8$~km~s$^{-1}$ & $85.2^{\circ} \pm 1.6^{\circ}$ & $0.60 \pm 0.10$ && $14.0 \pm 0.8$~km~s$^{-1}$ & $85.2^{\circ} \pm 1.2^{\circ}$ & $0.60 \pm 0.13$ \\
$3.00$~m~s$^{-1}$ && $14.0 \pm 0.8$~km~s$^{-1}$ & $84.8^{\circ} \pm 1.6^{\circ}$ & $0.60 \pm 0.17$ && $14.0 \pm 1.3$~km~s$^{-1}$ & $84.8^{\circ} \pm 1.6^{\circ}$ & $0.57 \pm 0.23$ \\
$5.00$~m~s$^{-1}$ && $14.0 \pm 1.5$~km~s$^{-1}$ & $86.6^{\circ} \pm 1.6^{\circ}$ & $0.63 \pm 0.28$ && $14.0 \pm 0.8$~km~s$^{-1}$ & $85.6^{\circ} \pm 2.4^{\circ}$ & $0.60 \pm 0.27$ \\
$7.00$~m~s$^{-1}$ && $14.0 \pm 1.5$~km~s$^{-1}$ & $86.0^{\circ} \pm 3.2^{\circ}$ & $0.67 \pm 0.42$ && $13.3 \pm 1.5$~km~s$^{-1}$ & $86.0^{\circ} \pm 3.2^{\circ}$ & $0.47 \pm 0.40$ \\
\hline
\end{tabular}
\end{table*}

\begin{table*}
\caption{Results of our fitting procedure for the WASP-7 RM simulations: 1) including different levels of instrumental noise, DR  and granulation; 2) including different levels of instrumental noise, DR, granulation and oscillation. The injected $\alpha$ was $0.3$ and the fit was performed accounting for differential rotation.}             
\label{resultsRMwithDRCBGRANOSC0.3}      
\centering                          
\begin{tabular}{ c c c c c c c c c }        
\hline\hline
Simulation &&  \multicolumn{3}{c}{DR + gran} && \multicolumn{3}{|c}{DR + gran + oscill} \\
\hline\hline 
noise && $v_*\sin i_*$ & $\lambda$ & $\alpha$ && $v_*\sin i_*$ & $\lambda$ & $\alpha$  \\  
\hline 
$0.50$~m~s$^{-1}$ && $13.8 \pm 0.3$~km~s$^{-1}$ & $85.4^{\circ} \pm 0.4^{\circ}$ & $0.23 \pm 0.03$ && $14.0 \pm 0.3$~km~s$^{-1}$ & $85.8^{\circ} \pm 0.4^{\circ}$ & $0.30 \pm 0.07$ \\
$1.00$~m~s$^{-1}$ && $13.8 \pm 0.3$~km~s$^{-1}$ & $85.0^{\circ} \pm 0.4^{\circ}$ & $0.23 \pm 0.10$ && $14.0 \pm 0.5$~km~s$^{-1}$ & $85.8^{\circ} \pm 0.8^{\circ}$ & $0.33 \pm 0.13$ \\
$2.00$~m~s$^{-1}$ && $14.0 \pm 0.5$~km~s$^{-1}$ & $84.4^{\circ} \pm 1.2^{\circ}$ & $0.30 \pm 0.10$ && $14.0 \pm 0.8$~km~s$^{-1}$ & $84.2^{\circ} \pm 1.2^{\circ}$ & $0.30 \pm 0.17$ \\
$3.00$~m~s$^{-1}$ && $13.5 \pm 0.8$~km~s$^{-1}$ & $83.6^{\circ} \pm 2.4^{\circ}$ & $0.13
^{+0.27}_{-0.13}$ && $14.0 \pm 1.3$~km~s$^{-1}$ & $84.4^{\circ} \pm 2.4^{\circ}$ & $0.30 \pm 0.23$ \\
$5.00$~m~s$^{-1}$ && $14.0 \pm 1.5$~km~s$^{-1}$ & $84.8^{\circ} \pm 3.2^{\circ}$ & $0.30 \pm 0.30$ && $14.0 \pm 1.3$~km~s$^{-1}$ & $84.2^{\circ} \pm 3.2^{\circ}$ & $0.30^{+0.33}_{-0.30}$  \\
$7.00$~m~s$^{-1}$ && $13.3 \pm 1.5$~km~s$^{-1}$ & $84.2^{\circ} \pm 3.2^{\circ}$ & $0.13^{+0.30}_{-0.13}$ && $14.0 \pm 2.3$~km~s$^{-1}$ & $84.2^{\circ} \pm 3.2^{\circ}$ & $0.13^{+0.40}_{-0.13}$  \\
\hline
                            
\end{tabular}
\end{table*}

\begin{table*}
\caption{Same as in Table \ref{resultsRMwithDRCBGRANOSC0.3} but for $\alpha = 0.6$.}             
\label{resultsRMwithDRCBGRANOSC0.6}      
\centering                          
\begin{tabular}{ c c c c c c c c c }        
\hline\hline
Simulation &&  \multicolumn{3}{c}{DR + gran} && \multicolumn{3}{|c}{DR + gran + oscill} \\
\hline\hline 
noise && $v_*\sin i_*$ & $\lambda$ & $\alpha$ && $v_*\sin i_*$ & $\lambda$ & $\alpha$  \\  
\hline 
$0.50$~m~s$^{-1}$ && $13.8 \pm 0.3$~km~s$^{-1}$ & $85.4^{\circ} \pm 0.4^{\circ}$ & $0.53 \pm 0.03$ && $14.0 \pm 0.3$~km~s$^{-1}$ & $85.8^{\circ} \pm 0.4^{\circ}$ & $0.60 \pm 0.07$ \\
$1.00$~m~s$^{-1}$ && $14.0 \pm 0.3$~km~s$^{-1}$ & $85.0^{\circ} \pm 0.8^{\circ}$ & $0.53 \pm 0.07$ && $14.0 \pm 0.5$~km~s$^{-1}$ & $85.6^{\circ} \pm 1.2^{\circ}$ & $0.63 \pm 0.10$ \\
$2.00$~m~s$^{-1}$ && $14.0 \pm 0.8$~km~s$^{-1}$ & $84.8^{\circ} \pm 1.2^{\circ}$ & $0.60 \pm 0.13$ && $14.0 \pm 0.8$~km~s$^{-1}$ & $84.2^{\circ} \pm 1.2^{\circ}$ & $0.60 \pm 0.17$\\
$3.00$~m~s$^{-1}$ && $14.0 \pm 0.5$~km~s$^{-1}$ & $84.8^{\circ} \pm 2.4^{\circ}$ & $0.60 \pm 0.13$ && $14.0 \pm 0.8$~km~s$^{-1}$ & $84.2^{\circ} \pm 1.6^{\circ}$ & $0.60 \pm 0.17$ \\
$5.00$~m~s$^{-1}$ && $14.0 \pm 1.5$~km~s$^{-1}$ & $84.8^{\circ} \pm 2.4^{\circ}$ & $0.60 \pm 0.30$ && $14.0 \pm 1.5$~km~s$^{-1}$ & $86.6^{\circ} \pm 2.4^{\circ}$ & $0.63 \pm 0.33$ \\
$7.00$~m~s$^{-1}$ && $14.0 \pm 1.5$~km~s$^{-1}$ & $84.2^{\circ} \pm 3.2^{\circ}$ & $0.47 \pm 0.30$ && $13.3 \pm 1.5$~km~s$^{-1}$ & $84.2^{\circ} \pm 3.6^{\circ}$ & $0.47 \pm 0.30$ \\
\hline
\end{tabular}
\end{table*}

\begin{table*}
\caption{Results of our fitting procedure for the WASP-7 RM simulations: 1) including instrumental noise ($ 2$~m~s$^{-1}$), DR  and different levels of granulation; 2) including instrumental noise ($2$~m~s$^{-1}$), DR and different levels of granulation (gran) and oscillation (oscill). The injected $\alpha$ was $0.3$ and the fit was performed accounting for differential rotation.}             
\label{resultsRMwithDRvariousgran0.3}      
\centering                          
\begin{tabular}{ c c c c c c c c c }        
\hline\hline
Simulation &&  \multicolumn{3}{c}{DR + gran} && \multicolumn{3}{|c}{DR + gran + oscill} \\
\hline\hline 
$gran$  && $v_*\sin i_*$ & $\lambda$ & $\alpha$ && $v_*\sin i_*$ & $\lambda$ & $\alpha$  \\  
\hline 
$0.33$ && $14.0 \pm 0.5$~km~s$^{-1}$ & $85.2^{\circ} \pm 1.2^{\circ}$ & $0.33 \pm 0.07$ && $14.0 \pm 0.5$~km~s$^{-1}$ & $85.2^{\circ} \pm 1.2^{\circ}$ & $0.30 \pm 0.13$ \\
$0.5$ && $14.0 \pm 0.5$~km~s$^{-1}$ & $85.2^{\circ} \pm 1.2^{\circ}$ & $0.30 \pm 0.10$ && $14.0 \pm 0.5$~km~s$^{-1}$ & $85.2^{\circ} \pm 1.2^{\circ}$ & $0.30 \pm 0.13$ \\
$1$ && $14.0 \pm 0.5$~km~s$^{-1}$ & $84.4^{\circ} \pm 1.2^{\circ}$ & $0.30 \pm 0.10$ && $14.0 \pm 0.8$~km~s$^{-1}$ & $84.2^{\circ} \pm 1.2^{\circ}$ & $0.30 \pm 0.17$ \\
$2$ && $13.5 \pm 0.5$~km~s$^{-1}$ & $83.0^{\circ} \pm 2.4^{\circ}$ & $0.13 \pm 0.13$ && $13.5 \pm 0.8$~km~s$^{-1}$ & $83.6^{\circ} \pm 1.2^{\circ}$ & $0.33 \pm 0.17$ \\
$3$ && $13.3 \pm 0.8$~km~s$^{-1}$ & $83.0^{\circ} \pm 2.0^{\circ}$ & $0.10 \pm 0.10$ && $14.0 \pm 0.8$~km~s$^{-1}$ & $83.6^{\circ} \pm 2.4^{\circ}$ & $0.33 \pm 0.20$ \\
\hline
                            
\end{tabular}
\end{table*}

\begin{table*}
\caption{Same as in Table \ref{resultsRMwithDRvariousgran0.3} but for $\alpha = 0.6$.}             
\label{resultsRMwithDRvariousgran0.6}      
\centering                          
\begin{tabular}{ c c c c c c c c c }        
\hline\hline
Simulation &&  \multicolumn{3}{c}{DR + gran} && \multicolumn{3}{|c}{DR + gran + oscill} \\
\hline\hline 
gran  && $v_*\sin i_*$ & $\lambda$ & $\alpha$ && $v_*\sin i_*$ & $\lambda$ & $\alpha$  \\  
\hline 
$0.33$ && $14.0 \pm 0.5$~km~s$^{-1}$ & $85.2^{\circ} \pm 1.6^{\circ}$ & $0.60 \pm 0.10$ && $14.0 \pm 0.5$~km~s$^{-1}$ & $85.2^{\circ} \pm 1.6^{\circ}$ & $0.60 \pm 0.13$ \\
$0.50$ && $14.0 \pm 0.5$~km~s$^{-1}$ & $85.2^{\circ} \pm 1.6^{\circ}$ & $0.60 \pm 0.13$ && $14.0 \pm 0.5$~km~s$^{-1}$ & $85.2^{\circ} \pm 1.6^{\circ}$ & $0.60 \pm 0.13$ \\
$1$ && $14.0 \pm 0.8$~km~s$^{-1}$ & $84.8^{\circ} \pm 1.2^{\circ}$ & $0.60 \pm 0.13$ && $14.0 \pm 0.8$~km~s$^{-1}$ & $84.2^{\circ} \pm 1.2^{\circ}$ & $0.60 \pm 0.17$\\
$2$ && $13.3 \pm 0.8$~km~s$^{-1}$ & $83.6^{\circ} \pm 1.6^{\circ}$ & $0.43 \pm 0.10$ && $14.0 \pm 0.8$~km~s$^{-1}$ & $84.2^{\circ} \pm 1.6^{\circ}$ & $0.63 \pm 0.20$ \\
$3$ && $13.3 \pm 0.8$~km~s$^{-1}$ & $82.4^{\circ} \pm 2.4^{\circ}$ & $0.40 \pm 0.10$ && $14.0 \pm 0.8$~km~s$^{-1}$ & $83.6^{\circ} \pm 2.4^{\circ}$ & $0.63 \pm 0.20$ \\
\hline
\end{tabular}
\end{table*}

\begin{table*}
\caption{Results of our fitting procedure  for the WASP-7 RM simulations which include a different injected FWHM, instrumental noise $\sigma = 2$~m~s$^{-1}$ and differential rotation ($\alpha = 0.3$ on the left side and $0.6$ on the right side). The fit was performed fixing FWHM $= 6.4$~km~s$^{-1}$}             
\label{resultsRMwithDRandFWHM0.3}      
\centering                          
\begin{tabular}{ c c c c c c c c c }        
\hline\hline
Simulation &&  \multicolumn{3}{c}{$\alpha = 0.3$} && \multicolumn{3}{c}{$\alpha = 0.6$} \\
\hline\hline
FWHM  && $v_*\sin i_*$ & $\lambda$ & $\alpha$ && $v_*\sin i_*$ & $\lambda$ & $\alpha$  \\  
\hline 
$6.0$~km~s$^{-1}$ && $14.3 \pm 0.3$~km~s$^{-1}$ & $86.0^{\circ} \pm 1.2^{\circ}$ & $0.33 \pm 0.10$ && $14.3 \pm 0.8$~km~s$^{-1}$ & $86.0^{\circ} \pm 1.2^{\circ}$ & $0.63 \pm 0.13$ \\
$6.8$~km~s$^{-1}$ && $14.0 \pm 0.8$~km~s$^{-1}$ & $85.2^{\circ} \pm 1.2^{\circ}$ & $0.33 \pm 0.13$ && $14.0 \pm 0.5$~km~s$^{-1}$ & $85.2^{\circ} \pm 2.4^{\circ}$ & $0.63 \pm 0.06$ \\
$7.2$~km~s$^{-1}$ && $13.8 \pm 0.3$~km~s$^{-1}$ & $85.2^{\circ} \pm 0.8^{\circ}$ & $0.30 \pm 0.13$ && $13.8 \pm 0.3$~km~s$^{-1}$ & $85.2^{\circ} \pm 1.2^{\circ}$ & $0.60 \pm 0.07$ \\
$7.6$~km~s$^{-1}$ && $13.8 \pm 0.8$~km~s$^{-1}$ & $85.2^{\circ} \pm 1.2^{\circ}$ & $0.33 \pm 0.10$ && $13.5 \pm 0.5$~km~s$^{-1}$ & $85.2^{\circ} \pm 1.2^{\circ}$ & $0.60 \pm 0.13$ \\
$8.0$~km~s$^{-1}$ && $13.5 \pm 0.3$~km~s$^{-1}$ & $85.2^{\circ} \pm 1.2^{\circ}$ & $0.33 \pm 0.10$ && $13.5 \pm 0.5$~km~s$^{-1}$ & $85.2^{\circ} \pm 1.6^{\circ}$ & $0.63 \pm 0.06$ \\
\hline                               
\end{tabular}
\end{table*}

\begin{table*}
\caption{Results of our fitting procedure  for the WASP-7 RM simulations which include a different limb darkening law, instrumental noise $\sigma = 2$~m~s$^{-1}$ and differential rotation ($\alpha = 0.3$). The fit was performed fixing $u_1 = 0.2$ and $u_2 = 0.3$.}
\label{resultsRMwithDRandLIMB0.3}      
\centering                          
\begin{tabular}{ c c c c c c c c c c }        
\hline\hline
\multicolumn{2}{c}{Simulation} &&  \multicolumn{3}{c}{$\alpha = 0.3$} && \multicolumn{3}{|c}{$\alpha = 0.6$} \\
\hline\hline
$u_1$ & $u_2$ && $v_*\sin i_*$ & $\lambda$ & $\alpha$ && $v_*\sin i_*$ & $\lambda$ & $\alpha$  \\  
\hline 
$0.25$ & $0.25$ && $14.0 \pm 0.5$~km~s$^{-1}$ & $86.0^{\circ} \pm 1.6^{\circ}$ & $0.30 \pm 0.13$ && $14.0 \pm 0.5$~km~s$^{-1}$ & $86.0^{\circ} \pm 1.6^{\circ}$ & $0.60 \pm 0.13$ \\
$0.15$ & $0.35$ && $14.0 \pm 0.5$~km~s$^{-1}$ & $86.0^{\circ} \pm 1.6^{\circ}$ & $0.30 \pm 0.13$ && $14.0 \pm 0.5$~km~s$^{-1}$ & $86.0^{\circ} \pm 1.6^{\circ}$ & $0.60 \pm 0.13$ \\
$0.25$ & $0.35$ && $14.5 \pm 0.5$~km~s$^{-1}$ & $86.0^{\circ} \pm 1.6^{\circ}$ & $0.43 \pm 0.13$ && $14.5 \pm 0.5$~km~s$^{-1}$ & $86.0^{\circ} \pm 1.6^{\circ}$ & $0.70 \pm 0.13$ \\

\hline                               
\end{tabular}
\end{table*}
\end{document}